\documentclass[12 pt,twoside,aps,prd,amsmath,amssymb,
tightenlines,showpacs,showkeys,eqsecnum]{revtex4}
\usepackage{epsfig}
\usepackage{mathptmx}
\usepackage{bm}
\usepackage{graphicx}
\renewcommand{\cal}{\mathcal}

\newcommand {\ve}{\varepsilon}
\newcommand {\pr}{\partial}
\newcommand {\cG}{\cal G}
\newcommand {\cD}{\cal D}
\newcommand {\G}{\Gamma}
\newcommand {\bg}{\bar \gamma}
\newcommand {\bp}{\bar \psi}

\newcommand {\p}{\psi}
\newcommand {\vf}{\varphi}

\numberwithin{equation}{section}

\def \myfigures #1#2#3#4#5#6#7#8
{\begin{figure}[ht]
    \begin{center}
        \includegraphics[width=#2 \textwidth]{#1.eps}
        \hfill
        \includegraphics[width=#6 \textwidth]{#5.eps}
        \parbox[t]{#4\textwidth}{\caption {#3}\label{#1}}
        \hfill
        \parbox[t]{#8\textwidth}{\caption {#7}\label{#5}}
    \end{center}
\end{figure} }

\def\myfigure #1#2#3#4
{\begin{figure}[ht]\begin{center}
\includegraphics[width=#2 \textwidth]{#1.eps}
\parbox[t]{#4\textwidth}{\caption{#3}\label{#1}}
\end{center}\end{figure}}

\begin{document}
\title{Bianchi type I cosmology with scalar and spinor fields}
\author{Bijan Saha and Todor Boyadjiev}
\affiliation{Laboratory of Information Technologies\\ 
Joint Institute for Nuclear Research, Dubna\\ 
141980 Dubna, Moscow region, Russia} 
\email{saha@thsun1.jinr.ru, todorlb@jinr.ru}

\begin{abstract}
We consider a system of interacting spinor and scalar fields in a 
gravitational field given by a Bianchi type-I cosmological model filled
with perfect fluid. The interacting term in the Lagrangian is chosen in 
the form of derivative coupling, i.e., 
${\cal L}_{\rm int} = \frac{\lambda}{2}\,\vf_{,\alpha}\vf^{,\alpha} F$,
with $F$ being a function of the invariants $I$ an $J$ constructed
from bilinear spinor forms $S$ and $P$. We consider the cases when
$F$ is the power or trigonometric functions of its arguments. 
Self-consistent solutions to the spinor, scalar and BI gravitational 
field equations are obtained. The problems of initial singularity and 
asymptotically isotropization process of the initially anisotropic
space-time are studied. 
It is also shown that the introduction of the Cosmological constant
($\Lambda$-term) in the Lagrangian generates oscillations of the BI model,
which is not the case in absence of $\Lambda$ term. Unlike the case
when spinor field nonlinearity is induced by self-action, in the case
in question, wehere nonlinearity is induced by the scalar field, there
exist regular solutions even without broken dominant energy condition.  
\end{abstract}

              

\maketitle

\bigskip


\section{Introduction}

Nonlinear generalization of classical field theory remains
one of the possible ways to overcome the difficulties of the 
theory which considers elementary particles as mathematical 
points. The gravitational field equation is nonlinear by 
nature and the field itself is universal and unscreenable.
These properties lead to definite physical interest for the proper
gravitational field to be considered. 
Nonlinear self-couplings of the spinor fields may arise as a consequence 
of the geometrical structure of the space-time and, more precisely, 
because of the existence of torsion. As early as 1938, Ivanenko
\cite{ivanenko1,ivanenko2,rodichev}
showed that a relativistic theory imposes in some cases a fourth-order
self-coupling. In 1950, Weyl~\cite{weyl} 
proved that, if the affine and the metric properties of the space-time
are taken as independent, the spinor field obeys either a linear equation
in a space with torsion or a nonlinear one in a Riemannian space. As
the self-action is of spin-spin type, it allows the assignment of a 
dynamical role to the spin and offers a clue about the origin of the
nonlinearities. A nonlinear spinor field, suggested by the symmetric 
coupling between nucleons, muons, and leptons, has been investigated
by Finkelstein et. al. \cite{finkel} in the classical approximation.  
Although the existence of spin-$1/2$ fermion is both theoretically and
experimentally undisputed, these are described by {\it quantum} spinor
fields. Possible justifications for the existence of classical spinors
has been addressed in \cite{greene}.

The present cosmology is based largely on Friedmann solutions
of the Einstein equations, which describe the completely uniform and
isotropic universe (``closed'' and ``open'' models)
The main feature of these solutions is their
non-stationarity. The idea of an expanding Universe, following from this
property, is confirmed by the astronomical observations and it is now safe
to assume that the isotropic model provides, in its general features,
an adequate description of the present state of the Universe.
Although the Universe seems homogeneous and isotropic at present, 
the large scale matter 
distribution in the observable universe, largely manifested in 
the form of discrete structures, does not exhibit homogeneity of 
a higher order. Recent space investigations detect anisotropy in 
the cosmic microwave background. In fact, 
the theoretical arguments~\cite{misner} and recent 
experimental data that support the existence of an anisotropic 
phase that approaches an isotropic one leads to consider the
models of universe with anisotropic back-ground. Zel'dovich was 
first to assume that the early isotropization of cosmological 
expanding process can take place as a result of quantum effect of 
particle creation near singularity~\cite{zel1}. This assumption was 
further justified by several authors~\cite{lu1,lu2,hu1}.

A Bianchi type-I (BI) Universe, being the straightforward generalization 
of the flat Friedmann-Robertson-Walker (FRW) Universe, is of particular
interest because it is one of the simplest models  
of an anisotropic Universe that describes a homogeneous and spatially flat
Universe. Unlike the FRW Universe which has the same 
scale factor for each of the three spatial directions, a BI Universe
has a different scale factor in each direction, thereby introducing an
anisotropy to the system. It moreover has the agreeable property that
near the singularity it behaves like a Kasner Universe, even in the 
presence of matter, and consequently falls within the general analysis
of the singularity given by Belinskii et al~\cite{belinskii}. 
Also in a Universe filled with matter for $p\,=\,\zeta\,\ve, \quad 
\zeta < 1$, it has been shown that any initial anisotropy in a BI
Universe quickly dies away and a BI Universe eventually evolves
into a FRW Universe~\cite{jacobs}. Since the present-day Universe is 
surprisingly isotropic, this feature of the BI Universe makes it a prime 
candidate for studying the possible effects of an anisotropy in the early 
Universe on present-day observations. 

It should be noted an important property of the isotropic model
is the presence of a singular point to time in its space-time metric
which means that the time is restricted. Is the
presence of a singular point an inherent property of the relativistic
cosmological models or it is only a consequence of specific
simplifying assumptions underlying these models? Motivated to answer this
question  self-consistent system of nonlinear spinor and BI gravitational
fields were studied in a series of papers 
\cite{sahajmp,sahaprd,sahagrg,sahal}. It should be mentioned that a spinor
field in a BI universe was also studied by Belinskii and Khalatnikov
\cite{belin}. Using Hamiltonian techniques, Henneaux studied class-A
Bianchi universes generated by a spinor source \cite{hen1,hen2}.

In this report we consider a self-consistent system of spinor, scalar 
and Bianchi type-I gravitation fields in presence of perfect fluid and
cosmological constant $\Lambda$. It should be noted that the inclusion
of the $\Lambda$ adds new dimension in the evolution of the universe.
Assuming that the $\Lambda$ term may be both possitive and negative,
it opens a much wider field of possibiliteis in the search for a 
singularity-free solutions to the field equations.
Apart from the papers written on this subject by us earlier, in this one 
beside the power type interaction term we consider a trigonometric one 
as well. In addition to asymptotic analysis, numerical analysis of
the corresponding nonlinear differential equations has been performed to
some extent as well. 

        \section{Derivation of Basic Equations}
In this section we derive the fundamental equations for the interacting 
spinor, scalar and gravitational fields from the action and write their
solutions in term of the voulume scale $\tau$ defined bellow \eqref{taudef}. 
We also derive the equation for $\tau$ which plays the central role here. 

   
We consider a system of nonlinear spinor, scalar and BI gravitational
field in presence of perfect fluid given by the action
\begin{equation}
{\cal S}(g; \p, \bp, \vf) = \int\, {\cal L} \sqrt{-g} d\Omega
\label{action}
\end{equation}
with 
\begin{equation} 
{\cal L} = {\cal L}_{\rm g} + {\cal L}_{\rm sp} + 
{\cal L}_{\rm sc} + {\cal L}_{\rm int} + {\cal L}_{\rm m}.
\label{lag} 
\end{equation} 
The gravitational part of the Lagrangian \eqref{lag} is given by a 
Bianchi type I (BI hereafter) space-time, whereas the remaining parts
are the usual spinor and scalar field Lagrangian with an interaction
between them and a perfect fluid as well.

             \subsection{Material field Lagrangian}

For a spinor field $\p$, symmetry between $\p$ and $\bp$ appears to
demand that one should choose the symmetrized Lagrangian~
\cite{kibble}.
Keep it in mind we choose the spinor field Lagrangian as 
\begin{equation} 
{\cal L}_{sp}=\frac{i}{2} 
\biggl[ \bp \gamma^{\mu} \nabla_{\mu} \p- \nabla_{\mu} \bp 
\gamma^{\mu} \p \biggr] - m\bp \p,  \label{lspin}
\end{equation} 
with $m$ being the spinor mass.

The mass-less scalar field Lagrangian is chosen to be
\begin{equation}
{\cal L}_{\rm sc} = \frac{1}{2} \vf_{,\alpha}\vf^{,\alpha}. 
\label{lsc}
\end{equation} 

The interaction between the spinor and scalar fields is given by the 
Lagrangian~\cite{sahagrg}
\begin{equation}
{\cal L}_{\rm int}= \frac{\lambda}{2}\,\vf_{,\alpha}\vf^{,\alpha} F,
\label{lint}
\end{equation}
with $\lambda$ being the coupling constant and $F$ is some 
arbitrary functions of invariants generated from the real bilinear 
forms of a spinor field with the form
\begin{subequations}
\label{bf}
\begin{eqnarray}
 S&=& \bar \psi \psi\qquad ({\rm scalar}),   \\                   
  P&=& i \bar \psi \gamma^5 \psi\qquad ({\rm pseudoscalar}), \\
 v^\mu &=& (\bar \psi \gamma^\mu \psi) \qquad ({\rm vector}),\\
 A^\mu &=&(\bar \psi \gamma^5 \gamma^\mu \psi)\qquad 
({\rm pseudovector}), \\
Q^{\mu\nu} &=&(\bar \psi \sigma^{\mu\nu} \psi)\qquad
({\rm antisymmetric\,\,\, tensor}),  
\end{eqnarray}
\end{subequations}
where $\sigma^{\mu\nu}\,=\,(i/2)[\gamma^\mu\gamma^\nu\,-\,
\gamma^\nu\gamma^\mu]$. 
Invariants, corresponding to the bilinear forms, are
\begin{subequations}
\label{invariants}
\begin{eqnarray}
I &=& S^2, \\
J &=& P^2, \\ 
I_v &=& v_\mu\,v^\mu\,=\,(\bar \psi \gamma^\mu \psi)\,g_{\mu\nu}
(\bar \psi \gamma^\nu \psi),\\ 
I_A &=& A_\mu\,A^\mu\,=\,(\bar \psi \gamma^5 \gamma^\mu \psi)\,
g_{\mu\nu}(\bar \psi \gamma^5 \gamma^\nu \psi), \\
I_Q &=& Q_{\mu\nu}\,Q^{\mu\nu}\,=\,(\bar \psi 
\sigma^{\mu\nu} \psi)\,g_{\mu\alpha}g_{\nu\beta}
(\bar \psi \sigma^{\alpha\beta} \psi). 
\end{eqnarray}
\end{subequations}

According to the Pauli-Fierz theorem \cite{Ber} among the five 
invariants only $I$ and $J$ are independent as all other can be 
expressed by them:
$I_V = - I_A = I + J$ and $I_Q = I - J.$ Therefore, we choose 
$F = F(I, J)$, thus claiming that it describes the nonlinearity in 
the most general of its form \cite{sahaprd}. Note that setting
$\lambda = 0$ is \eqref{lint} we come to the case with minimal coupling. 
        
The term ${\cal L}_{\rm m}$ describes the Lagrangian density of 
perfect fluid.

             \subsection{The gravitational field}
As a gravitational field we consider the Bianchi type I (BI) cosmological
model. It is the simplest model of anisotropic universe that describes
a homogeneous and spatially flat space-time and if filled with perfect
fluid with the equation of state $p = \zeta \ve, \quad \zeta < 1$, it
eventually evolves into a FRW universe \cite{jacobs}. The isotropy of 
present-day universe makes BI model a prime candidate for studying the
possible effects of an anisotropy in the early universe on modern-day
data observations. In view of what has been mentioned above we choose 
the gravitational part of the Lagrangian \eqref{lag} in the form
\begin{equation}
{\cal L}_{\rm g} = \frac{R}{2\kappa},
\label{lgrav}
\end{equation}
where $R$ is the scalar curvature, $\kappa = 8 \pi G$  
being the Einstein's gravitational constant. The gravitational field in 
our case is given by a Bianchi type I (BI) metric  
\begin{equation} 
ds^2 = dt^2 - a^2 dx^2 - b^2 dy^2 - c^2 dz^2, 
\label{BI}
\end{equation}
with $a,\, b,\, c$ being the functions of time $t$ only. Here the speed of
light is taken to be unity.

The metric \eqref{BI} has the following non-trivial Christoffel symbols
\begin{eqnarray}
\G^{1}_{10} &=& \frac{\dot a}{a}, \quad  \G^{2}_{20} = \frac{\dot b}{b},
\quad \G^{3}_{30} = \frac{\dot c}{c} \nonumber \\
\label{chrsym}\\
\G^{0}_{11} &=& a {\dot a}, \quad
\G^{0}_{22} = b {\dot b}, \quad \G^{0}_{33} = c {\dot c}. \nonumber
\end{eqnarray}
The nontrivial components of the Ricci tensors are
\begin{subequations}
\label{ricten}
\begin{eqnarray}
R_{0}^{0} &=& - \Bigl(\frac{\ddot a}{a}+\frac{\ddot b}{b}+\frac{\ddot c}{c}
\Bigr), \label{rt00} \\
R_{1}^{1} &=& - \Bigl[\frac{\ddot a}{a} + \frac{\dot a}{a}\Bigl(
\frac{\dot b}{b} + \frac{\ddot c}{c}\Bigr)\Bigr], \label{rt11} \\
R_{2}^{2} &=& - \Bigl[\frac{\ddot b}{b} + \frac{\dot b}{b}\Bigl(
\frac{\dot c}{c} + \frac{\ddot a}{a}\Bigr)\Bigr], \label{rt22}\\
R_{3}^{3} &=& - \Bigl[\frac{\ddot c}{c} + \frac{\dot c}{c}\Bigl(
\frac{\dot a}{a} + \frac{\ddot b}{b}\Bigr)\Bigr]. \label{rt33}
\end{eqnarray}
\end{subequations}
From \eqref{ricten} one finds the following Ricci scalar for the BI universe 
\begin{equation}
R = - 2\Bigl(\frac{\ddot a}{a} + \frac{\ddot b}{b} + \frac{\ddot c}{c} +
\frac{\dot a}{a}\frac{\dot b}{b} + \frac{\dot b}{b}\frac{\dot c}{c}+
\frac{\dot c}{c}\frac{\dot a}{a}\Bigr).
\label{ricsc}
\end{equation}
The non-trivial components of Riemann tensors in this case read
\begin{eqnarray}
R^{01}_{\,\,\,\,01} &=& \frac{\ddot a}{a}, \quad 
R^{02}_{\,\,\,\,02} = \frac{\ddot b}{b}, \quad 
R^{03}_{\,\,\,\,03} = \frac{\ddot c}{c}, \nonumber \\
\label{riemann} \\
R^{12}_{\,\,\,\,12} &=& -\frac{\dot a}{a}\frac{\dot b}{b}, \quad 
R^{23}_{\,\,\,\,23} = -\frac{\dot b}{b}\frac{\dot c}{c}, \quad 
R^{31}_{\,\,\,\,31} = -\frac{\dot c}{c}\frac{\dot a}{a}. \nonumber
\end{eqnarray}
Now having all the non-trivial components of Ricci and Riemann tensors,
one can easily write the invariants of gravitational field which we need 
to study the space-time singularity. We return to this study at the end 
of this section.

  \subsection{Field equations}
Let us now write the field equations corresponding to the action
\eqref{action}. 

Variation of \eqref{action} with respect to spinor field $\psi\,(\bp)$
gives spinor field equations
\begin{subequations}
\label{speq}
\begin{eqnarray}
i\gamma^\mu \nabla_\mu \psi - m \psi + {\cD} \psi + 
{\cG} i \gamma^5 \psi &=&0, \label{speq1} \\
i \nabla_\mu \bp \gamma^\mu +  m \bp - {\cD} \bp - 
{\cG} i \bp \gamma^5 &=& 0, \label{speq2}
\end{eqnarray}
\end{subequations}
where we denote
$$ {\cD} =  \lambda S \vf_{,\alpha}\vf^{,\alpha} \frac{\pr F}{\pr I}, 
\quad
{\cG} =  \lambda P \vf_{,\alpha}\vf^{,\alpha} \frac{\pr F}{\pr J}.$$
Since the nonlinearity in the spinor field equations is generated by the
interacting scalar field, the Eqs. \eqref{speq} can be viewed as
spinor field equations with induced nonlinearity.

Varying \eqref{action} with respect to scalar field
yields the following scalar field equation
\begin{equation}
\frac{1}{\sqrt{-g}} \frac{\pr}{\pr x^\nu} \Bigl(\sqrt{-g} g^{\nu\mu}
(1 + \lambda F) \vf_{,\mu}\Bigr) 
= 0. \label{scfe}
\end{equation}

Finally, variation of \eqref{action} with respect to metric tensor 
$g_{\mu\nu}$ gives the Einstein's field equation in account of the
$\Lambda$-term has the form
\begin{equation}
R_{\mu}^{\nu} - \frac{1}{2} \delta_{\mu}^{\nu} = \kappa T_{\mu}^{\nu}
- \delta_{\mu}^{\nu} \Lambda.
\label{Ein}
\end{equation}
In view of \eqref{ricten} and \eqref{ricsc} for the BI space-time \eqref{BI}
we rewrite the Eq. \eqref{Ein} as
\begin{subequations}
\label{BID}
\begin{eqnarray}
\frac{\ddot b}{b} +\frac{\ddot c}{c} + \frac{\dot b}{b}\frac{\dot 
c}{c}&=&  \kappa T_{1}^{1} -\Lambda,\label{11}\\
\frac{\ddot c}{c} +\frac{\ddot a}{a} + \frac{\dot c}{c}\frac{\dot 
a}{a}&=&  \kappa T_{2}^{2} - \Lambda,\label{22}\\
\frac{\ddot a}{a} +\frac{\ddot b}{b} + \frac{\dot a}{a}\frac{\dot 
b}{b}&=&  \kappa T_{3}^{3} - \Lambda,\label{33}\\
\frac{\dot a}{a}\frac{\dot b}{b} +\frac{\dot b}{b}\frac{\dot c}{c} 
+\frac{\dot c}{c}\frac{\dot a}{a}&=&  \kappa T_{0}^{0} - \Lambda,
\label{00}
\end{eqnarray}
\end{subequations}
where over dot means differentiation with respect to $t$ 
and $T_{\nu}^{\mu}$ is the energy-momentum tensor
of the material field given by
\begin{equation}
T_{\mu}^{\nu} = T_{{\rm sp}\,\mu}^{\,\,\,\nu} + T_{{\rm sc}\,\mu}^{\,\,\,\nu}
+ T_{{\rm int}\,\mu}^{\,\,\,\nu} + T_{{\rm m}\,\mu}^{\,\,\,\nu}.
\label{tem}
\end{equation}
Here $T_{{\rm sp}\,\mu}^{\,\,\,\nu}$ is the energy-momentum tensor of 
the spinor field defined by   
\begin{equation}
T_{{\rm sp}\,\mu}^{\,\,\,\rho}=\frac{i}{4} g^{\rho\nu} \biggl(\bp \gamma_\mu 
\nabla_\nu \psi + \bp \gamma_\nu \nabla_\mu \psi - \nabla_\mu \bar 
\psi \gamma_\nu \psi - \nabla_\nu \bp \gamma_\mu \psi \biggr) \,-
\delta_{\mu}^{\rho}{\cal L}_{\rm sp}.
\label{temsp}
\end{equation}
The term  ${\cal L}_{sp}$ with respect to \eqref{speq} takes the form
\begin{equation}
{\cal L}_{sp} = -\bigl({\cD} S + {\cG} P\bigr).
\label{lsp}
\end{equation}
The energy-momentum tensor of the scalar field 
$T_{{\rm sc}\,\mu}^{\,\,\,\nu}$ is given by 
\begin{equation}
T_{{\rm sc}\,\mu}^{\,\,\,\nu}= \vf_{,\mu}\vf^{,\nu}
- \delta_{\mu}^{\nu}{\cal L}_{\rm sc}.
\label{temsc}
\end{equation}
For the interaction field we find
\begin{equation}
T_{{\rm int}\,\mu}^{\,\,\,\nu}= \lambda F \vf_{,\mu}\vf^{,\nu}
 -  \delta_{\mu}^{\nu} {\cal L}_{\rm int}.
\label{temint}
\end{equation}
$T_{\mu\,{\rm (m)}}^{\nu}$ is the energy-momentum tensor of a perfect fluid. 
For a universe filled with perfect fluid, in a comoving system of reference
such that $u^\mu = (1,\,0,\,0,\,0)$ we have
\begin{equation}  
T_{\mu\,{\rm (m)}}^{\nu} = (p + \ve) u_\mu u^\nu - \delta_\mu^\nu p = 
(\ve,\,-p,\,-p,\,-p),
\end{equation}
where energy of the perfect fluid 
$\ve$ is related to its' pressure $p$ by the equation 
of state $p\,=\,\zeta\,\ve$.  Here $\zeta$ varies between the
interval $0\,\le\, \zeta\,\le\,1$, whereas $\zeta\,=\,0$ describes
the dust Universe, $\zeta\,=\,\frac{1}{3}$ presents radiation Universe,
$\frac{1}{3}\,<\,\zeta\,<\,1$ ascribes hard Universe and $\zeta\,=\,1$
corresponds to the stiff matter. 

In the Eqs. \eqref{speq} and \eqref{temsp} $\nabla_\mu$ is the covariant
derivatives acting on a spinor field as ~\cite{zhelnorovich,brill}
\begin{equation}
\label{cvd}
\nabla_\mu \psi = \frac{\partial \psi}{\partial x^\mu} -\G_\mu \psi, \quad
\nabla_\mu \bp = \frac{\partial \bp}{\partial x^\mu} + \bp \G_\mu, 
\end{equation}
where $\G_\mu$ are the Fock-Ivanenko spinor connection coefficients
defined by
\begin{equation}
\G_\mu = \frac{1}{4} \gamma^\sigma \Bigl(\G_{\mu \sigma}^{\nu} \gamma_{\nu}
- \partial_{\mu} \gamma_{\sigma}\Bigr).  
\label{fock}
\end{equation}  
For the metric \eqref{BI} one has the following components
of the spinor connection coefficients 
\begin{eqnarray} 
\G_0 = 0, \quad 
\G_1 = \frac{1}{2}\dot a(t) \bg^1 \bg^0, \quad
\G_2 = \frac{1}{2}\dot b(t) \bg^2 \bg^0, \quad 
\G_3 = \frac{1}{2}\dot c(t) \bg^3 \bg^0. 
\label{ficc}
\end{eqnarray}
The Dirac matrices $\gamma^\mu(x)$ of curved space-time are 
connected with those of Minkowski one as follows:
$$ \gamma^0=\bg^0,\quad \gamma^1 =\bg^1/a,
\quad \gamma^2=\bg^2 /b,\quad \gamma^3 =\bg^3 /c$$
with 
\begin{eqnarray}
\bg^0\,=\,\left(\begin{array}{cc}I&0\\0&-I\end{array}\right), \quad
\bg^i\,=\,\left(\begin{array}{cc}0&\sigma^i\\
-\sigma^i&0\end{array}\right), \quad
\gamma^5 = \bg^5&=&\left(\begin{array}{cc}0&-I\\
-I&0\end{array}\right),\nonumber
\end{eqnarray}
where $\sigma_i$ are the Pauli matrices:
\begin{eqnarray}
\sigma^1\,=\,\left(\begin{array}{cc}0&1\\1&0\end{array}\right), 
\quad
\sigma^2\,=\,\left(\begin{array}{cc}0&-i\\i&0\end{array}\right), 
\quad
\sigma^3\,=\,\left(\begin{array}{cc}1&0\\0&-1\end{array}\right).
\nonumber
\end{eqnarray}
Note that the $\bg$ and the $\sigma$ matrices obey the following 
properties:
\begin{eqnarray}
\bg^i \bg^j + \bg^j \bg^i = 2 \eta^{ij},\quad i,j = 0,1,2,3 
\nonumber\\
\bg^i \bg^5 + \bg^5 \bg^i = 0, \quad (\bg^5)^2 = I, 
\quad i=0,1,2,3 \nonumber\\
\sigma^j \sigma^k = \delta_{jk} + i \varepsilon_{jkl} \sigma^l, 
\quad j,k,l = 1,2,3 \nonumber
\end{eqnarray}
where $\eta_{ij} = \{1,-1,-1,-1\}$ is the diagonal matrix, 
$\delta_{jk}$ is the Kronekar symbol and $\varepsilon_{jkl}$ 
is the totally antisymmetric matrix with $\varepsilon_{123} = +1$.
 
We study the space-independent solutions to the spinor 
and scalar field equations \eqref{speq} and \eqref{scfe} so that 
$\psi=\psi(t)$ and $\vf = \vf(t)$.
Defining
\begin{equation}
\tau = a b c = \sqrt{-g}
\label{taudef}
\end{equation}
from \eqref{scfe} for the scalar field  we have
\begin{equation}
\vf = C \int \frac{dt}{\tau (1 + \lambda F)}, \quad C = {\rm const.}
\label{sfsol}
\end{equation}

The spinor field equation \eqref{speq1} in account of \eqref{cvd} and 
\eqref{ficc} takes the form
\begin{equation} i\bg^0 
\biggl(\frac{\partial}{\partial t} +\frac{\dot \tau}{2 \tau} \biggr) \psi 
- m \psi + {\cD}\psi + {\cG} i \gamma^5 \psi = 0.
\label{spq}
\end{equation} 
Setting $V_j(t) = \sqrt{\tau} \psi_j(t), \quad j=1,2,3,4,$ from 
\eqref{spq} one deduces the following system of equations:  
\begin{subequations}
\label{V}
\begin{eqnarray} 
\dot{V}_{1} + i (m - {\cD}) V_{1} - {\cG} V_{3} &=& 0, \\
\dot{V}_{2} + i (m - {\cD}) V_{2} - {\cG} V_{4} &=& 0, \\
\dot{V}_{3} - i (m - {\cD}) V_{3} + {\cG} V_{1} &=& 0, \\
\dot{V}_{4} - i (m - {\cD}) V_{4} + {\cG} V_{2} &=& 0. 
\end{eqnarray} 
\end{subequations}

From \eqref{speq1} we also write the equations for the invariants
$S,\quad P$ and $A = \bp \bg^5 \bg^0 \psi$
\begin{subequations}
\label{inv}
\begin{eqnarray}
{\dot S_0} - 2 {\cG}\, A_0 &=& 0, \label{S0}\\
{\dot P_0} - 2 (m - {\cD})\, A_0 &=& 0, \label{P0}\\
{\dot A_0} + 2 (m - {\cD})\, P_0 + 2 {\cG} S_0 &=& 0, \label{A0} 
\end{eqnarray}
\end{subequations}
where $S_0 = \tau S, \quad P_0 = \tau P$, and $ A_0 = \tau A.$
The Eqn. \eqref{inv} leads to the following relation
\begin{equation}
S^2 + P^2 + A^2 =  C^2/ \tau^2, \qquad C^2 = {\rm const.}
\label{inv1}
\end{equation}

Giving the concrete form of $F$ from \eqref{V} one writes
the components of the spinor functions in explicitly and
using the solutions obtained one can write the components of
spinor current:
\begin{equation}
j^\mu = \bp \gamma^\mu \psi.
\label{spincur}
\end{equation}
The component $j^0$ 
\begin{equation}
j^0 = \frac{1}{\tau}
\bigl[V_{1}^{*} V_{1} + V_{2}^{*} V_{2} + V_{3}^{*} V_{3}
+ V_{4}^{*} V_{4}\bigr], 
\end{equation}
defines the charge density of spinor field 
that has the following chronometric-invariant form 
\begin{equation}
\varrho = (j_0\cdot j^0)^{1/2}. 
\label{rho}
\end{equation}
The total charge of spinor field is defined as
\begin{equation}
Q = \int\limits_{-\infty}^{\infty} \varrho \sqrt{-^3 g} dx dy dz =
   \varrho \tau {\cal V},
\label{charge}
\end{equation}
where ${\cal V}$ is the volume. From the spin tensor
\begin{equation}
S^{\mu\nu,\epsilon} = \frac{1}{4}\bp \bigl\{\gamma^\epsilon
\sigma^{\mu\nu}+\sigma^{\mu\nu}\gamma^\epsilon\bigr\} \psi.
\label{spin}
\end{equation}
one finds chronometric invariant spin tensor 
\begin{equation}
S_{{\rm ch}}^{ij,0} = \bigl(S_{ij,0} S^{ij,0}\bigr)^{1/2},
\label{chij}
\end{equation} 
and the projection of the spin vector on $k$ axis 
\begin{equation}
S_k = \int\limits_{-\infty}^{\infty} S_{{\rm ch}}^{ij,0} 
\sqrt{-^3 g} dx dy dz = S_{{\rm ch}}^{ij,0} \tau V. 
\label{proj}
\end{equation} 

Let us now solve the Einstein equations. To do it we first write the 
expressions for the components of the energy-momentum tensor explicitly:
\begin{eqnarray}
\label{total}
T_{0}^{0} &=& mS + \frac{C^2}{2\tau^2 (1+\lambda F)} + \ve, \nonumber\\
\\
T_{1}^{1} &=& T_{2}^{2} = T_{3}^{3} =
{\cD} S + {\cG} P - \frac{C^2}{2\tau^2 (1+\lambda F)} 
- p.\nonumber 
\end{eqnarray}
In account of \eqref{total} subtracting \eqref{11} from \eqref{22},
one finds the following relation between $a$ and $b$
\begin{equation}
\frac{a}{b}= D_1 \mbox{exp} \biggl(X_1 \int \frac{dt}{\tau}\biggr).  
\label{ab}
\end{equation}
Analogically, one finds
\begin{equation} 
\frac{a}{c}= D_2 \mbox{exp} \biggl(X_2 \int \frac{dt}{\tau}\biggr), \quad 
\frac{b}{c}= D_3 \mbox{exp} \biggl(X_3 \int \frac{dt}{\tau}\biggr).  
\label{ac}
\end{equation}
Here $D_1,\,D_2,\,D_3,\,X_1,\, X_2, X_3 $ are integration constants, obeying
\begin{eqnarray}
D_1 D_2 D_3 = 1, \quad X_1 + X_2 + X_3 = 0.
\label{intcon}
\end{eqnarray}

In view of \eqref{intcon} from \eqref{ab} and \eqref{ac} 
we write the metric functions explicitly~\cite{sahaprd}
\begin{subequations}
\label{abc}
\begin{eqnarray} 
a(t) &=& 
(D_{1}^{2}D_{3})^{1/3}\tau^{1/3}\mbox{exp}\biggl[\frac{2 X_1 + X_3 
}{3} \int\,\frac{dt}{\tau (t)} \biggr], \label{a} \\
b(t) &=& 
(D_{1}^{-1}D_{3})^{1/3}\tau^{1/3}\mbox{exp}\biggl[-\frac{X_1 - X_3 
}{3} \int\,\frac{dt}{\tau (t)} \biggr], \label{b} \\
c(t) &=& 
(D_{1}D_{3}^{2})^{-1/3}\tau^{1/3}\mbox{exp}\biggl[-\frac{X_1 + 2 X_3 
}{3} \int\,\frac{dt}{\tau (t)} \biggr].  \label{c}
\end{eqnarray}
\end{subequations}
As one sees from \eqref{a}, \eqref{b} and \eqref{c}, for $\tau = t^n$
with $n > 1$ the exponent tends to unity at large $t$, and the 
anisotropic model becomes isotropic one. 

Further we will investigate the existence of singularity (singular point)
of the gravitational case, which can be done by investigating the
invariant characteristics of the space-time. In general relativity 
these invariants are composed from the curvature tensor and the 
metric one. In a 4D Riemann space-time there are 14 independent invariants. 
Instead of analyzing all 14 invariants, one can confine
this study only in 3, namely the scalar curvature $I_1 = R$, 
$I_2 = R_{\mu\nu}^R{\mu\nu}$, and the Kretschmann scalar
$I_3 = R_{\alpha\beta\mu\nu}R^{\alpha\beta\mu\nu}$.
At any regular space-time point, these three invariants 
$I_1,\,I_2,\,I_3$ should be finite. Let us rewrite these invariants
in detail.

For the Bianchi I metric one finds the scalar curvature 
\begin{eqnarray}
I_1 = R = - 2\Bigl(\frac{\ddot a}{a}+\frac{\ddot b}{b}+\frac{\ddot c}{c}+
\frac{\dot a}{a}\frac{\dot b}{b} + \frac{\dot b}{b}\frac{\dot c}{c}+
\frac{\dot c}{c}\frac{\dot a}{a}\Bigr).
\label{SC}
\end{eqnarray}
Since the Ricci tensor for the BI metric is diagonal, the 
invariant $I_2 = R_{\mu\nu}R^{\mu\nu} \equiv R_{\mu}^{\nu} R_{\nu}^{\mu}$
is a sum of squares of diagonal components of Ricci tensor, i.e.,
\begin{equation}
I_2 = \Bigl[\bigl(R_{0}^{0}\bigr)^2  + \bigl(R_{1}^{1}\bigr)^2 +
\bigl(R_{2}^{2}\bigr)^2 + \bigl(R_{3}^{3}\bigr)^2 \Bigr],
\end{equation}
with the components of the Ricci tensor being given by \eqref{ricten}.

Analogically, for the Kretschmann scalar in this case we have 
$I_3 = R_{\,\,\,\,\,\,\,\alpha\beta}^{\mu\nu}
R_{\,\,\,\,\,\,\,\mu\nu}^{\alpha\beta}$,
a sum of squared components of all nontrivial 
$R_{\,\,\,\,\,\,\,\mu\nu}^{\mu\nu}$, which in view of \eqref{riemann}
can be written as
\begin{eqnarray}
I_3 &=& 4 \Biggl[  
 \Bigl(R_{\,\,\,\,\,\,01}^{01}\Bigr)^2  
+ \Bigl(R_{\,\,\,\,\,\,02}^{02}\Bigr)^2
+ \Bigl(R_{\,\,\,\,\,\,03}^{03}\Bigr)^2
+ \Bigl(R_{\,\,\,\,\,\,12}^{12}\Bigr)^2
+ \Bigl(R_{\,\,\,\,\,\,23}^{23}\Bigr)^2
+ \Bigl(R_{\,\,\,\,\,\,31}^{31}\Bigr)^2\Biggr] \nonumber\\
&=& 4\Bigl[\Bigl(\frac{\ddot a}{a}\Bigr)^2 + 
\Bigl(\frac{\ddot b}{b}\Bigr)^2+\Bigl(\frac{\ddot c}{c}\Bigr)^2
+ \Bigl(\frac{\dot a}{a}\frac{\dot b}{b}\Bigr)^2 + 
\Bigl(\frac{\dot b}{b}\frac{\dot c}{c}\Bigr)^2 +
\Bigl(\frac{\dot c}{c}\frac{\dot a}{a}\Bigr)^2\Bigr].
\label{Kretsch}
\end{eqnarray}
Let us now express the forgoing invariants in terms of $\tau$.
From Eqs. \eqref{abc} we have 
\begin{subequations}
\label{sing}
\begin{eqnarray}
a_i &=& A_i \tau^{1/3} {\rm exp} \Biggl((Y_i/3) \int \tau^{-1} dt\Biggr),  
\\
\frac{{\dot a}_i}{a_i} &=& \frac{Y_i+1}{3}\frac{1}{\tau} \quad (i=1,2,3,),\\
\frac{{\ddot a}_i}{a_i} &=& \frac{(Y_i+1)(Y_i-2)}{9}\frac{1}{\tau^2},
\end{eqnarray}
\end{subequations}
i.e., the metric functions $a, b, c$ and their derivatives are in 
functional dependence with $\tau$. From Eqs. \eqref{sing} one can easily
verify that
$$I_1 \propto \frac{1}{\tau^2},\quad
I_2 \propto \frac{1}{\tau^4},\quad I_3 \propto \frac{1}{\tau^4}.$$
Thus we see that at any space-time point, where $\tau = 0$ the invariants 
$I_1,\,I_2,\,I_3$, as well as the scalar and spinor fields
become infinity, hence the space-time becomes singular at this point.  

In what follows, we write the equation for $\tau$ and study it in details.  

Summation of Einstein equations \eqref{11}, \eqref{22}, \eqref{33} and 
\eqref{00} multiplied by 3 gives
\begin{equation}
\frac{\ddot 
\tau}{\tau}= \frac{3}{2}\kappa \Bigl(mS + {\cD} S + {\cG} P + \ve -p
\Bigr) - 3 \Lambda. 
\label{dtau1}
\end{equation} 
For the right-hand-side of \eqref{dtau1} to be a function
of $\tau$ only, the solution to this equation is well-known~\cite{kamke}.

Let us demand the energy-momentum to be conserved, i.e.,
\begin{equation}
T_{\mu;\nu}^{\nu} = T_{\mu,\nu}^{\nu} + \G_{\rho\nu}^{\nu} T_{\mu}^{\rho}
- \G_{\mu\nu}^{\rho} T_{\rho}^{\nu} = 0,
\end{equation}
which in our case has the form
\begin{equation}
\frac{1}{\tau}\bigl(\tau T_0^0\bigr)^{\cdot} - \frac{\dot a}{a} T_1^1
-\frac{\dot b}{b} T_2^2  - \frac{\dot c}{c} T_3^3 = 0.
\label{emcon}
\end{equation}
In account of the equation of state $p = \zeta \ve$ and 
$$(m -{\cD}) \dot{S}_0 - {\cG} \dot{P}_0 = 0$$
which follows from \eqref{inv}, after a little manipulation from
\eqref{emcon} we obtain  
\begin{equation}
\ve = {\ve_0}/{\tau^{1+\zeta}},\quad 
p = {\zeta \ve_0}/{\tau^{1+\zeta}}.
\label{vep}
\end{equation}
In view of \eqref{vep} the Eq. \eqref{dtau1} can be written as
\begin{equation}
\frac{\ddot \tau}{\tau}= \frac{3}{2}\kappa \Bigl(mS + {\cD} S + {\cG} P + 
(1 - \zeta) \ve_0/ \tau^{1+\zeta}\Bigr) - 3 \Lambda. 
\label{dtau2}
\end{equation}
As it was mentioned earlier, we consider $F$ as a function of $I$, $J$ or 
$I\pm J$. In the section to follow we study the Eq. \eqref{dtau2} in details.

        \section{Exact solutions and numerical analysis} 

In the preceding section we solved the spinor, scalar and gravitational
field equations and wrote the solutions in terms of volume-scale $\tau$.
It was also mentioned that for the right hand side of the Eq. \eqref{dtau2}
to be the function of $\tau$, this equation is quadratically integrable.
In what follows, we explicitly write the solutions for corresponding
equations given some concrete form of $\tau$.  

        \subsection{Exact solutions}
Here we consider the cases with minimal coupling and with $F$ being the
function of either $I$ or $J$ (with zero mass). In this subsection we simply
write the solutions to the spinor field  equations explicitly and
present the solution for $\tau$ in quadrature. 

    \subsubsection{Minimally coupled scalar and spinor fields}
Let us first consider the case with minimal coupling when the scalar and
the spinor fields interact through gravitational one. In this case
from \eqref{inv} one finds $S = C_0/\tau$. Scalar field and the components 
of the spinor field in this case have the following explicit form
\begin{equation}
\vf = C \int \frac{dt}{\tau},
\label{sclin}
\end{equation}
\begin{eqnarray} 
\psi_1(t) &=& \frac{C_1}{\sqrt{\tau}} e^{-imt}, \quad
\psi_2(t) = \frac{C_2}{\sqrt{\tau}} e^{-imt},  \nonumber\\
\label{splin}\\
\psi_3(t) &=& \frac{C_3}{\sqrt{\tau}} e^{imt}, \quad
\psi_4(t) = \frac{C_4}{\sqrt{\tau}} e^{imt},
\nonumber
\end{eqnarray} 
with the integration constants $C_j$ satisfying
$C_0 = C_1^2 + C_2^2 - C_3^2 - C_4^2.$ 

Eq. \eqref{dtau2} in this case takes the form
\begin{equation}
{\ddot \tau}= \frac{3}{2}\kappa \Bigl(m C_0 +  \ve_0 (1 - \zeta)/
\tau^{\zeta} \Bigr) - 3 \Lambda \tau, 
\label{dtaulin}
\end{equation}
with the solution
\begin{equation}
\int \frac{d \tau}{\sqrt{\kappa\Bigl(m C_0 \tau 
+ \ve_0 \tau^{1 - \zeta}  
\Bigr) - \Lambda \tau^2 + E }} \,= \, \sqrt{3}\, t.
\label{quadralin}
\end{equation}
Here $E$ is the constant of integration. Let us note that being the
volume-scale $\tau$ cannot be negative. On the other hand the radical
in \eqref{quadralin} should be positive. This fact leads to the 
conclusion that for a positive $\Lambda$ the value of $\tau$ is bound
from above giving rise to an oscillatory mode of expansion of the BI 
space-time.

    \subsubsection{Case with $F = F(I)$}

Here we consider the interacting system of scalar and spinor field 
with the interaction given by 
${\cal L}_{int} = (\lambda/2) \vf_\mu \vf^\mu F(I).$  
As in the case with minimal coupling from \eqref{S0} one finds
\begin{equation}
S = \frac{C_0}{\tau}, \quad C_0 = {\rm const.}
\label{stau}
\end{equation}
For  components of spinor field we find~\cite{sahaprd}
\begin{eqnarray} 
\psi_1(t) &=& \frac{C_1}{\sqrt{\tau}} e^{-i\beta}, \quad
\psi_2(t) = \frac{C_2}{\sqrt{\tau}} e^{-i\beta},  \nonumber\\
\label{spef}\\
\psi_3(t) &=& \frac{C_3}{\sqrt{\tau}} e^{i\beta}, \quad
\psi_4(t) = \frac{C_4}{\sqrt{\tau}} e^{i\beta},
\nonumber
\end{eqnarray} 
with $C_i$ being the integration constants and
are related to $C_0$ as 
$C_0 = C_{1}^{2} + C_{2}^{2} - C_{3}^{2} - C_{4}^{2}.$ Here
$\beta = \int(m - {\cD}) dt$.

For the components of the spin current from \eqref{spincur} we find
\begin{eqnarray}
j^0 &=& \frac{1}{\tau}
\bigl[C_{1}^{2} + C_{2}^{2} + C_{3}^{2} + C_{4}^{2}\bigr],\quad
j^1 = \frac{2}{a\tau}
\bigl[C_{1} C_{4} + C_{2} C_{3}\bigr] {\rm cos}(2\beta),
\nonumber \\
j^2 &=& \frac{2}{b\tau}
\bigl[C_{1} C_{4} - C_{2} C_{3}\bigr] {\rm sin}(2\beta),\quad
j^3 = \frac{2}{c\tau}
\bigl[C_{1} C_{3} - C_{2} C_{4}\bigr] {\rm cos}(2\beta), \nonumber
\end{eqnarray}
whereas, for the projection of spin vectors on the $X$, $Y$ and $Z$
axis we find
\begin{eqnarray}
S^{23,0} = \frac{C_1 C_2 + C_3 C_4}{b c\tau},\quad
S^{31,0} = 0,\quad
S^{12,0} = \frac{C_1^2 - C_2^2 + C_3^2 - C_4^2}{2ab\tau}. \nonumber
\end{eqnarray}
Total charge of the system in a volume $\cal{V}$ in this case is
\begin{equation}
Q = [C_1^2 + C_{2}^{2} + C_{3}^{2} + C_{4}^{2}] \cal{V}.
\end{equation}
Thus, for $\tau \ne 0$ the components of spin current and
the projection of spin vectors are singularity-free and the total 
charge of the system in a finite volume is always finite.

The equation for determining $\tau$ in this case has the form
\begin{equation}
{\ddot \tau}= \frac{3}{2}\kappa \Bigl(m C_0 + {\cD} C_0 +  \ve_0 (1 - \zeta)/
\tau^{\zeta}  \Bigr) - 3 \Lambda \tau. 
\label{dtaui}
\end{equation} 
Recalling that ${\cD} = \lambda C_0 C^2 F_I /\tau^3 (1 + \lambda F(I))^2$
the solution to Eq. \eqref{dtaui} can be written in quadrature
\begin{equation}
\int \frac{d \tau}{\sqrt{\kappa\Bigl(m C_0 \tau + C^2/2(1 + \lambda F)
+ \ve_0 \tau^{1 - \zeta}\Bigr) - \Lambda \tau^2 + E }} \,= \, \sqrt{3}\, t,
\label{quadrai}
\end{equation}
with $E$ being some integration constant. Given the explicit form of $F(I)$
for different $\Lambda$ we have find different mode of expansion. We study
this case in details numerically in the subsection to follow.

    \subsubsection{Case with $F = F(J)$}

Finally we consider the interacting system of scalar and spinor field 
with the interaction given by 
${\cal L}_{int} = (\lambda/2) \vf_\mu \vf^\mu F(J).$  
In the case considered we assume the spinor field to be massless.
It gives ${\cD} = 0$. Note that, in the unified 
nonlinear spinor theory of Heisenberg, the massive term remains 
absent, and according to Heisenberg, the particle mass should be 
obtained as a result of quantization of spinor prematter~
\cite{massless}. In the nonlinear generalization of classical field 
equations, the massive term does not possess the significance that 
it possesses in the linear one, as it by no means defines total 
energy (or mass) of the nonlinear field system. Thus without losing 
the generality we can consider mass-less spinor field putting $m\,=\,0.$ 
Then from \eqref{P0} one gets
\begin{equation}
P = D_0/\tau, \quad D_0 = {\rm const.}
\label{ptau}
\end{equation}
In this case the spinor field components take the form
\begin{eqnarray}
\psi_1 &=&\frac{1}{\sqrt{\tau}} \bigl(D_1 e^{i \sigma} + 
iD_3 e^{-i\sigma}\bigr), \quad
\psi_2 =\frac{1}{\sqrt{\tau}} \bigl(D_2 e^{i \sigma} + 
iD_4 e^{-i\sigma}\bigr), \nonumber \\
\label{J}\\
\psi_3 &=&\frac{1}{\sqrt{\tau}} \bigl(iD_1 e^{i \sigma} + 
D_3 e^{-i \sigma}\bigr),\quad
\psi_4 =\frac{1}{\sqrt{\tau}} \bigl(iD_2 e^{i \sigma} + 
D_4 e^{-i\sigma}\bigr). \nonumber
\end{eqnarray} 
The integration constants $D_i$
are connected to $D_0$ by
$D_0=2\,(D_{1}^{2} + D_{2}^{2} - D_{3}^{2} -D_{4}^{2}).$
Here we set $\sigma = \int {\cG} dt$. 

For the components of the spin current from \eqref{spincur} we find
\begin{eqnarray}
j^0 &=& \frac{2}{\tau}
\bigl[D_{1}^{2} + D_{2}^{2} + D_{3}^{2} + D_{4}^{2}\bigr],\quad
j^1 = \frac{4}{a\tau}
\bigl[D_{2} D_{3} + D_{1} D_{4}\bigr] {\rm cos}(2 \sigma), \nonumber\\
j^2 &=& \frac{4}{b\tau}
\bigl[D_{2} D_{3} - D_{1} D_{4}\bigr] {\rm sin}(2 \sigma),\quad
j^3 = \frac{4}{c\tau}
\bigl[D_{1} D_{3} - D_{2} D_{4}\bigr] {\rm cos}(2 \sigma), \nonumber
\end{eqnarray}
whereas, for the projection of spin vectors on the $X$, $Y$ and $Z$
axis we find
\begin{eqnarray}
S^{23,0} = \frac{2(D_{1} D_{2} + D_{3} D_{4})}{b c\tau},\quad
S^{31,0} = 0,\quad
S^{12,0} = \frac{D_{1}^{2} - D_{2}^{2} + D_{3}^{2} - D_{4}^{2}}{2ab\tau}
\nonumber
\end{eqnarray}

For $\tau$ in this case we have
\begin{equation}
{\ddot \tau}= \frac{3}{2}\kappa \Bigl({\cG} C_0 +  \ve_0 (1 - \zeta)/
\tau^{\zeta} \Bigr) - 3 \Lambda \tau. 
\label{dtauj}
\end{equation} 
In view of \eqref{ptau}, ${\cG}$ in this case takes the form analogical
to that taken by ${\cD}$ in previous case with $F_I$ replaced by $F_J$.
Then solution to Eq. \eqref{dtauj} we write in quadrature as
\begin{equation}
\int \frac{d \tau}{\sqrt{\kappa\Bigl(C^2/2(1 + \lambda F)
+ \ve_0 \tau^{1 - \zeta}\Bigr) - \Lambda \tau^2 + E }}\,=\,\sqrt{3}\,t,
\label{quadraj}
\end{equation}
Depending on the form of $F$ and $\Lambda$ we have different mode of 
expansion of BI universe as in previous case. In what follows we numerically
study the aforementioned cases.

 \subsection{Numerical experiments}
In this subsection we numerically solve the Eq. \eqref{dtaui} for some
different choice of $F$. As it was mentioned earlier, setting $\lambda = 0$
in \eqref{dtaui} we come to the case with minimal coupling given by 
\eqref{dtaulin}, whereas, assuming $m = 0$ we get \eqref{dtauj}. Let us
first rewrite Eq. \eqref{dtaui}
\begin{equation}
    {\ddot \tau} = \mathcal F(\tau, p)\,,
\label{dtaunui}
\end{equation}
where we denote
\begin{equation}
   \mathcal F \equiv  \frac{3}{2}\kappa \Bigl(m C_0 + {\cD} C_0 + 
 \ve_0 (1 - \zeta)/\tau^{\zeta}  \Bigr) - 3 \Lambda \tau,
\label{pot}
\end{equation}
and 
$p \equiv \{\kappa, \lambda, m, C_0, C, \varepsilon_0, \zeta, \Lambda \}$ is 
the set of the parameters. Since in the examples we consider $F = F(S)$,
let us rewrite ${\cD}$ in terms of $S$.
In account of $ S = C_0/\tau$ for ${\cD}$ we have
 $${\cD} = \lambda C^2 F_S /2\tau^2 (1 + \lambda F(S))^2.$$ 
From mechanical point of view the Eqn. \eqref{dtaunui} can be interpreted 
as an equation of motion of a single particle with unit mass under the 
force $\mathcal F(\tau,p)$. Then the following first integral exists 
\cite{ll_76}
\begin{equation} 
    \dot \tau = \sqrt{2[E - \mathcal U(\tau,p)]}\,.
\label{1stint}
\end{equation}
Here $E$ is the integration constant and 
$$\mathcal{U} \equiv - \frac{3}{2} \Bigl[\kappa \Bigl(m C_0 \tau +
C^2/2(1 + \lambda F)+\ve_0 \tau^{-\zeta}\Bigr) - \Lambda \tau^2\Bigr]\,,$$
is the potential of the force $\mathcal F$. We note that the radical 
expression must be non-negative. The zeroes of this expression, which 
depend on all the problem parameters $p$ define the boundaries of the 
possible rates of changes of $\tau(t)$. In what follows we numerically
analyze \eqref{dtaunui} and \eqref{pot} for different choice of $F(I)$
as well as for different problem parameters $p$.

        \subsubsection{$F = S^n$}
Let us first choose $F$ to be a power function of $S$(or $I$), setting
$F = S^n$. 
In this case setting $C_0 = 1$ and $C = 1$ we  rewrite ${\mathcal F}$ as
\begin{equation}
 {\mathcal F} = \frac{3\kappa}{2}\Bigl(m + \frac{\lambda\, n\, \tau^{n-1}}
{2\, (\lambda  + \tau^n)^2} + \ve_0 \frac{(1-\zeta)}{\tau^\zeta} \Bigr) - 3 
\Lambda\, \tau, \label{nueq}
\end{equation}
with the potential
\begin{equation}
  \mathcal{U} = - \frac{3}{2} \left\{\kappa \left [m\, \tau - \frac{\lambda}
{2\,(\lambda + \tau^n) }+\ve_0 \tau^{1-\zeta}
\right] - \Lambda \tau^2 \right\}. 
\label{quads}
\end{equation}
Note that the nonnegativity of the radical in \eqref{1stint} in view of 
\eqref{quads} imposes restriction on $\tau$ from above in case of 
$\Lambda > 0$. It means that in case of $\Lambda > 0$ the value of 
$\tau$ runs between $0$ and some $\tau_{\rm max}$, where $\tau_{\rm max}$ 
is the maximum value of $\tau$ for the given value of $p$. 
This equation has been studied for different values of parameters $p$. 
Here we demonstrate the evolution of $\tau$ for different choice of 
$\tau_0$ for fixed ``energy'' $E$ and vise versa.
  
As the first example we consider massive spinor field with $m = 1$.
Other parameters are chosen in the following way: 
coupling constant $\lambda = 0.1$, power of nonlinearity $n = 4$, 
and cosmological constant $\Lambda = 1/3$. We also choose $\zeta = 0.5$
describing a hard universe.

\vskip 1 cm
\myfigures{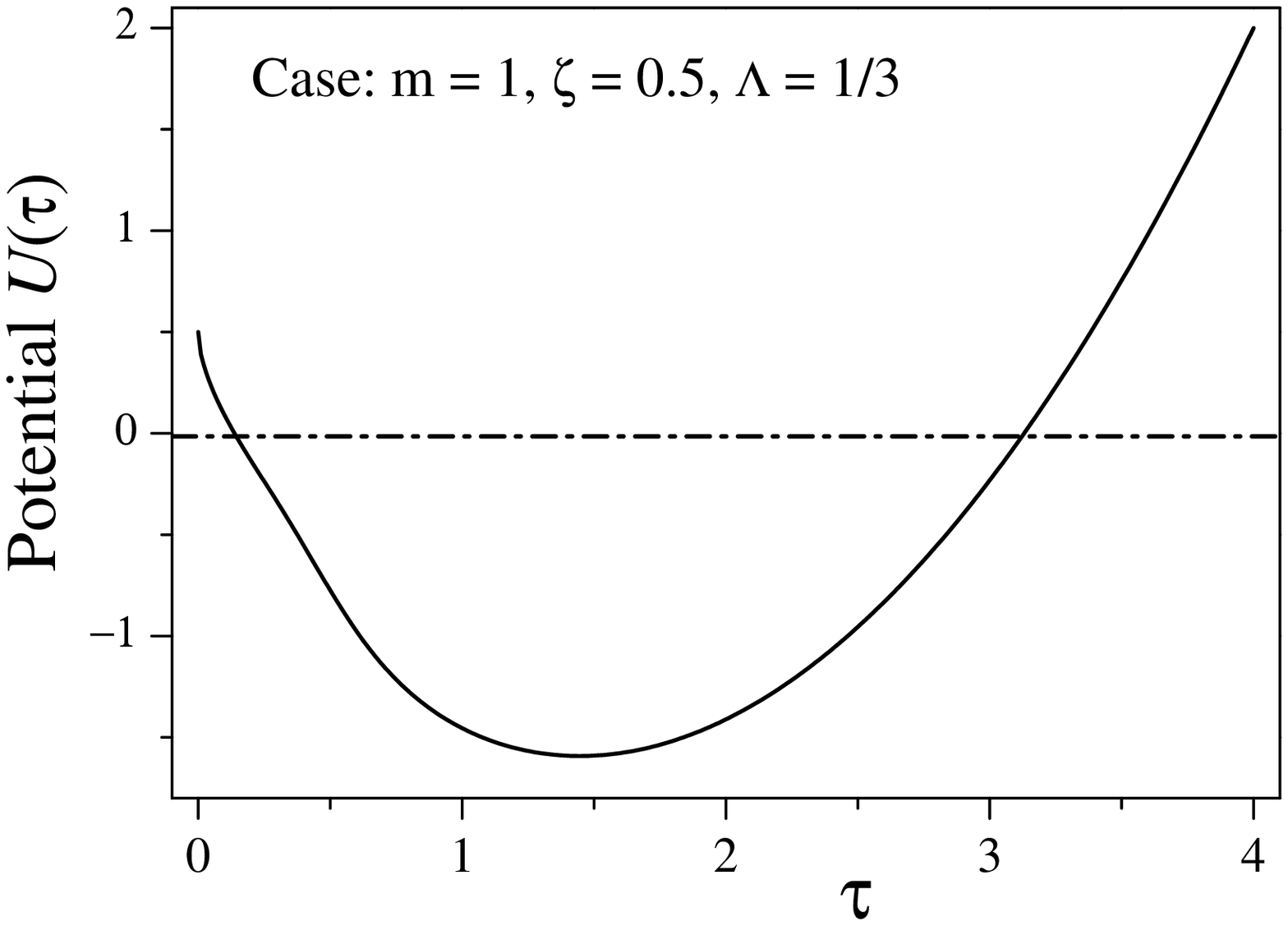}{0.45}{Perspective view of the potential 
$\mathcal U(\tau)$ with BI space-time being filled perfect fluid 
describing hard universe}{0.45}{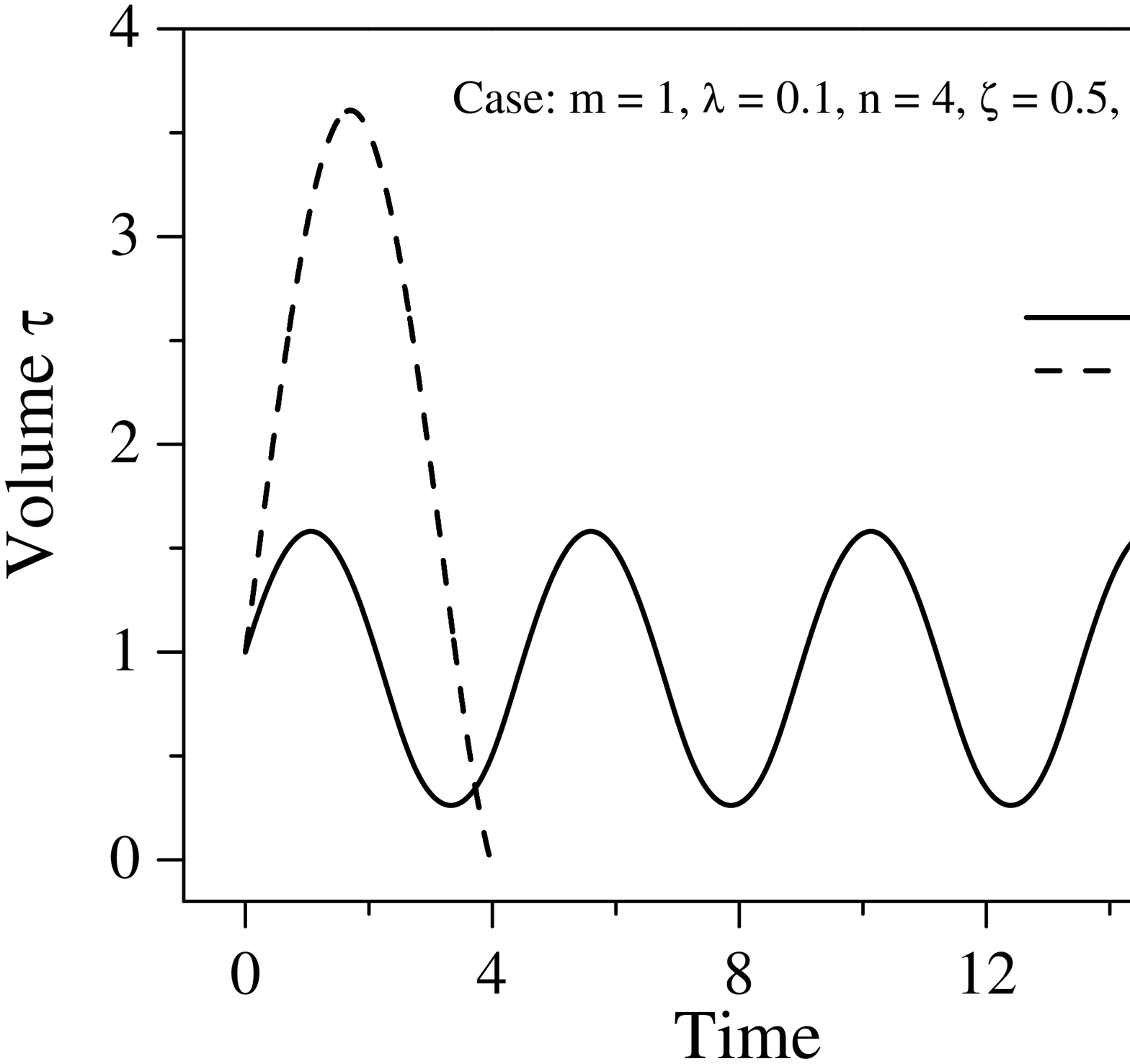}{0.45}{Evolution of the BI 
space-time corresponding to the potential given in Fig.~\ref{pot6} 
for different choice of $E$.}{0.45}

In Fig. \ref{pot6} we plot corresponding potential $\mathcal U(\tau)$ 
multiplied by the factor $2/3$. As is seen from Fig.~\ref{pot6} and 
Fig.~\ref{tau6}, choosing the integration constant $E$ we may obtain two 
different types of solutions. For $E > 0.5$ solutions are non-periodic, 
whereas for $E_{\rm min} < E \le 0.5$ the evolution of the universe is
oscillatory.

As a second example we consider the massless spinor field. Other parameters
of the problem are left unaltered with the exception of $\zeta$. Here we
choose $\zeta = 1$ describing stiff matter. It should be noted that
this particular choice of $\zeta$ gives rise to a local maximum. 
This results in two types of solutions for a single choice of $E$.

\vskip 1 cm
\myfigures{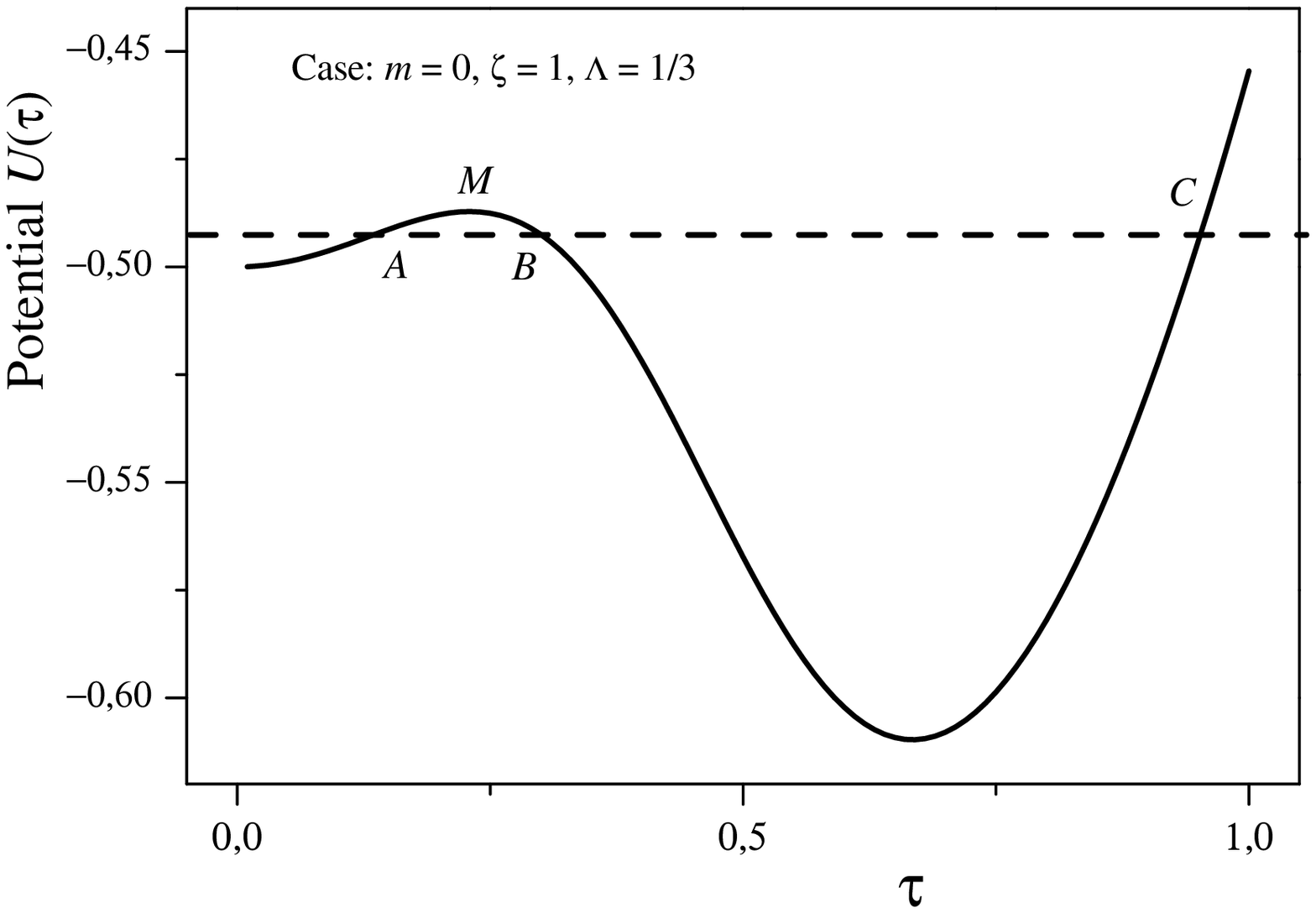}{0.45}{Perspective view of the potential 
$\mathcal U(\tau)$ with BI space-time being filled with stiff 
matter}{0.45}{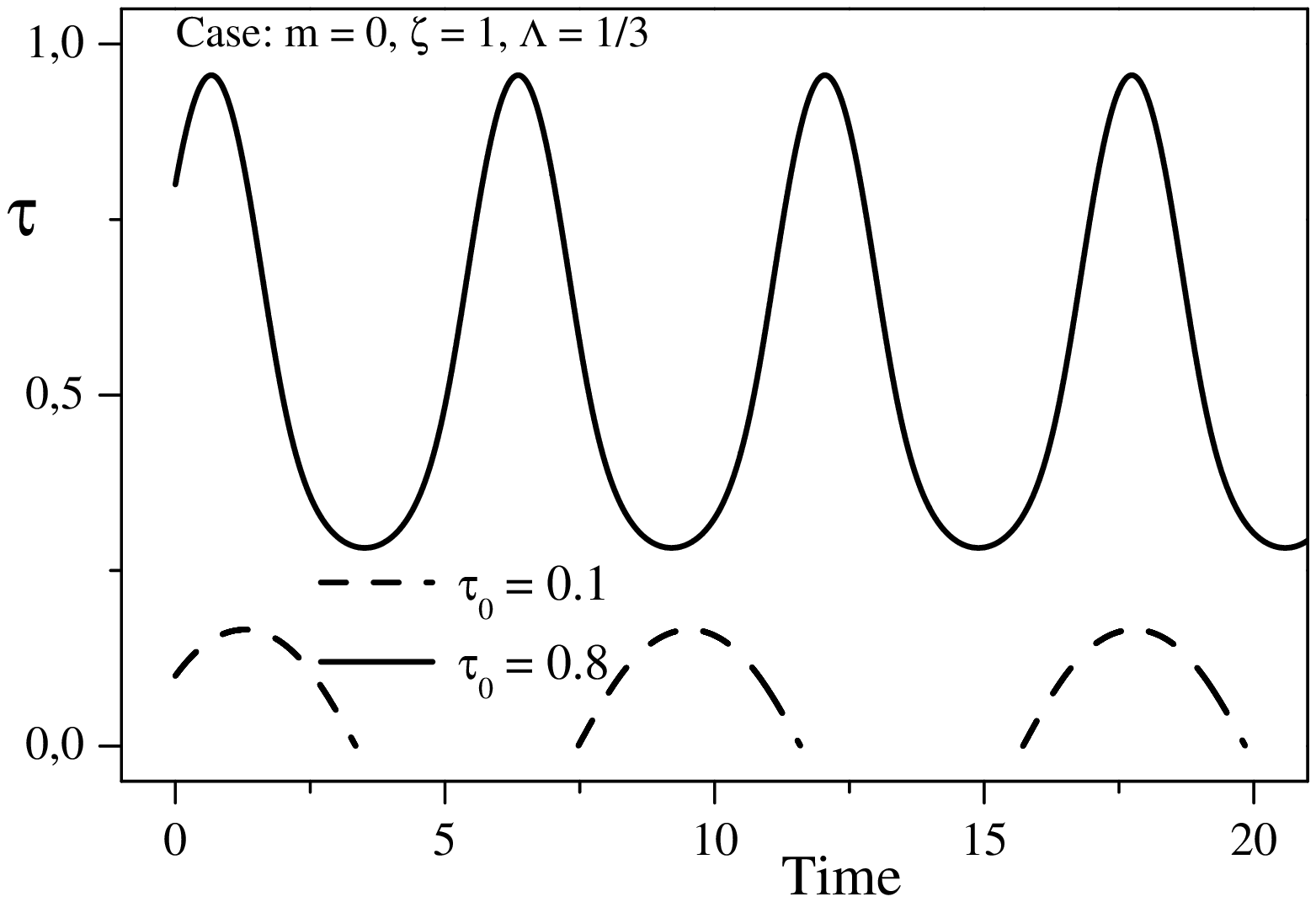}{0.43}{Evolution of the BI space-time corresponding 
to the potential given in Fig.~\ref{pot3} in case of massless spinor field 
for different choice of $\tau_0$ with $E \in (-0.5,\, M)$.}{0.43}

As one sees from Fig.~\ref{pot3}, for $E > M$ there exists only non-periodic
solutions, whereas, for $E_{\rm min} < E < -0.5$ the solutions are always
oscillatory. For $E \in (-0.5,\, M)$ there exits two types of solutions
depending on the choice of $\tau_0$. In Fig.~\ref{tau3} we plot the 
evolution of $\tau$ for $E \in (-0.5,\, M)$. As is seen, for 
$\tau_0 \in (0,\, A)$ we have periodical solution, but due to the fact
that $\tau$ is non-negative, the physical solutions happen to be
semi-periodic. For $\tau_0 \in (B,\, C)$ we again have oscillatory mode
of the evolution of $\tau$. This two region is separated by a no-solution
zone $(A,\,B)$. 

Let us also consider the case with $\Lambda < 0$. For a negative $\Lambda$,
as well as in absence of the $\Lambda$-term
the evolution of $\tau$ is always exponential as it is seen in
Fig.~\ref{Taucase2}. In this case the initial anisotropy of the BI
space-time quickly dies away and the universe becomes isotropic one.

\vskip 1 cm
\myfigure{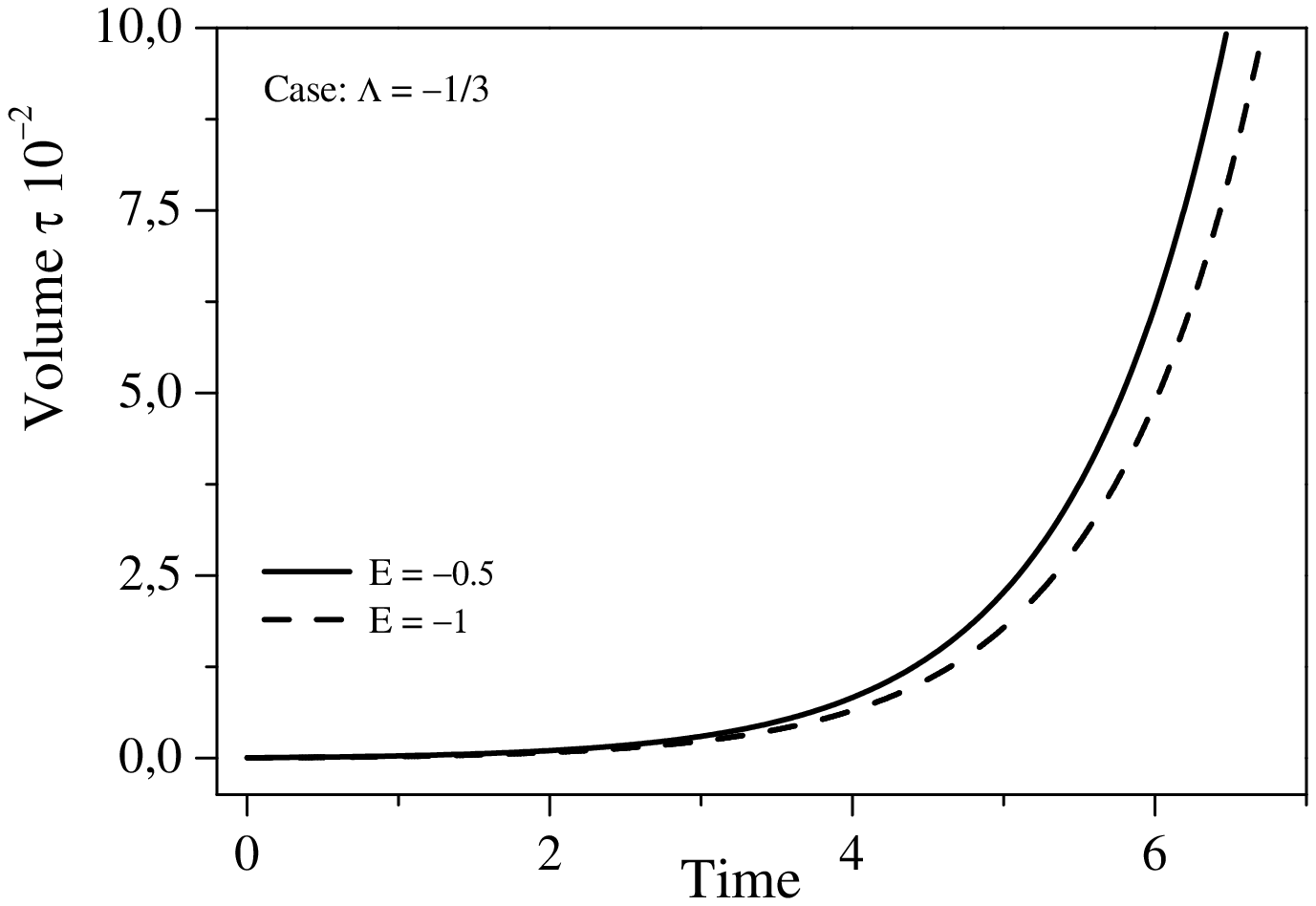}{0.60}{Perspective view of $\tau$ for a negative
$\Lambda$. As one sees, evolution of the universe in this case takes
exponential character and the initial anisotropy of the BI space-time
quickly dies away.}{0.60}

Let us analyze the dominant energy condition in the Hawking-Penrose 
theorem \cite{hawking}. For a BI universe  
the dominant energy condition can be written in the form \cite{sahaprd}
\begin{subequations}
\label{dec2}
\begin{eqnarray}
T_{0}^{0} &\ge& T_{1}^{1} a^2 + T_{2}^{2} b^2 + T_{3}^{3} c^2,\\
T_{0}^{0} &\ge& T_{1}^{1} a^2, \\
T_{0}^{0} &\ge& T_{2}^{2} b^2, \\
T_{0}^{0} &\ge& T_{3}^{3} c^2.
\end{eqnarray}
\end{subequations}
It was shown that for the spinor field with self-coupling the regular 
solutions can be obtained only by violating energy-dominanat condition 
\cite{sahaprd}. Regular solution that does not violate this condition 
was found only for linear spinor case by means of a positive $\Lambda$ 
term. To analyze this condition for the spinor field with induced 
nonlinearity let us rewrite the components of energy-momentum tensor. 
For energy density in this case we have
\begin{equation}
T_0^0 = \frac{m C_0}{\tau}+\frac{C^2\tau^{n-2}}{2 (\tau^n + \lambda C_0^n)}
+\frac{\ve_0}{\tau^{1+\zeta}}.
\label{eden}
\end{equation}
As one sees from \eqref{eden} for any positive value of $\tau$ energy density
is always positive definite. As  $\tau \to 0$, $T_0^0 \to \infty$, whereas
$T_0^0$ decreases as $\tau$ increases. For the pressure components we have
\begin{equation}
T_1^1 = T_2^2 = T_3^3 =
\frac{C^2\tau^{n-2}}{2 (\tau^n + \lambda C_0^n)^2} 
\bigl[\lambda C_0^n (n-1) - \tau^n\bigr] -\frac{\zeta \ve_0}{\tau^{1+\zeta}}.
\label{pden}
\end{equation}
The second term in \eqref{pden} is always positive, it means $T_1^1$ 
has a greater value when BI universe is filled with dust, i.e., when 
$\zeta = 0$. To investigate the dominant energy condition we study the 
pressure term (since $T_1^1 = T_2^2 = T_3^3$, hereafter we mention it as
$T_1^1$) in details. For simplicity we set $C = 1$ and $C_0 = 1$. It is
clear from \eqref{pden} that if
\begin{equation}
\tau^n > \lambda (n - 1),
\label{eneq}
\end{equation}
we have $T_1^1 < 0$. In this case the dominant energy condition remains 
unbroken. From \eqref{eneq} we see, for $\lambda = 0$ the foregoing
inequality holds for any $\tau > 0$. It means like the linear spinor field,
the system of minimally coupled scalar and spinor fields possesses regular
solutions without violating the dominant energy condition. For an 
interacting system this condition holds for any negative $n$ with a positive
$\lambda$ and vice versa. Let us now see what happens when both $n$ and
$\lambda$ are positive (negative). Note that the coupling constant $\lambda$ 
may take any value. The magnitude of $\lambda$ defines the strength of 
interaction. 

It is clear that for a large value of $\tau$ the inequality \eqref{eneq}
is likely to be held for any reasonable value of $\lambda$. For $\tau$
close to zero it is unlikely to be held, but in that case, as it was 
mentioned earlier, $T_0^0$ tends to $\infty$, so the dominant energy 
condition at this region remains safe. Finally we see what happens to it
when $\tau$ is close to unity. $T_0^0$ at this point is reasonably small,
whereas for a relatively large $n$ we have a situation when $T_1^1$ 
dominates $T_0^0$. Relative behavior of $T_0^0$ and $T_1^1$ has been
shown in Fig.~\ref{en_mom1}. Thus we conclude that in case of interacting
spinor and scalar fields it is possible to construct regular solutions
without violating dominant energy condition of Hawking-Penrose theorem.

\vskip 1 cm
\myfigures{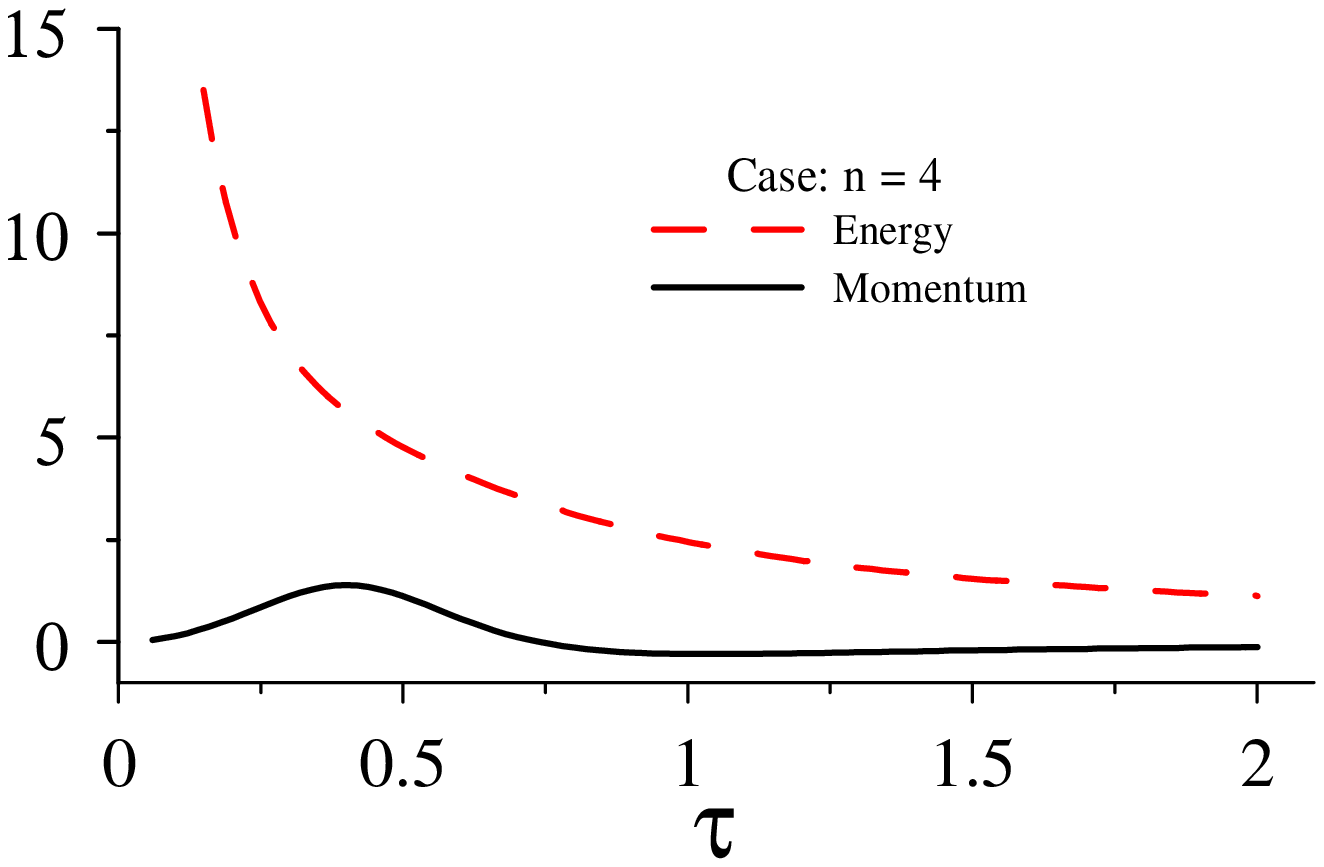}{0.45}{Perspective view of $T_0^0$ and $T_1^1$ for a positive
$n$. As one sees for a small value of $n$ it is possible to construct
a regular solution without violating dominant energy condition.}{0.45}
{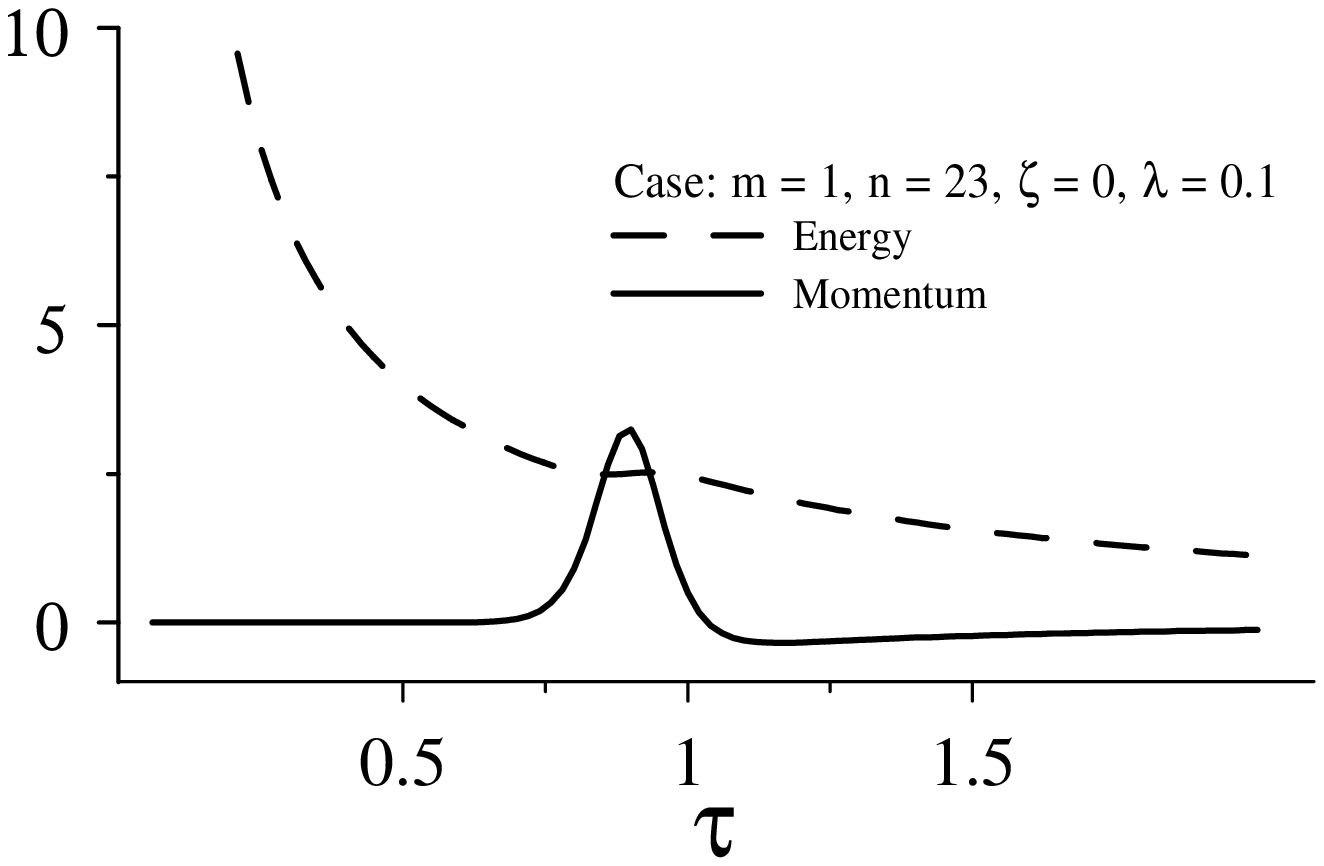}{0.43}
{For a large $n$ there exists some value of $\tau$ where the pressure
component prevails energy. In this case the dominant energy condition
breaks down.}{0.43}

        \subsubsection{$F = \sin\,S$}

Let us now consider the case with $F = \sin\,S$. In this case 
for ${\mathcal F}$ we have
\begin{equation}
 {\mathcal F} = \frac{3\kappa}{2}\Bigl(m + \frac{\lambda\, \cos\,S}
{2\,\tau^2 (\lambda  + \sin\,S)^2} + \ve_0 \frac{(1-\zeta)}
{\tau^\zeta} \Bigr) - 3 \Lambda\, \tau, \label{nueqsin}
\end{equation}
with the potential
\begin{equation}
  \mathcal{U} = - \frac{3}{2} \left\{\kappa \left [m\, \tau + \frac{1}
{2\,(1 + \lambda\,\sin\,S)} + \ve_0 \tau^{1-\zeta} \right] - 
\Lambda \tau^2  \right\}. \label{quadsin}
\end{equation}
It should be noted that unlike the case with $F$ being a power function
of $\tau$ where the nonlinearity appears in the region with large value
of $\tau$, in the case under consideration, a number of interesting properties
emerge in the region where $0 < \tau < 1$, namely, in the vicinity of the
singular point $\tau = 0$. A perspective view of the potential 
 $\mathcal U(\tau)$ is given in the Figs. \ref{c01} and \ref{c01a}. Here
we choose the problem parameters as follows: $\kappa = 2/3$, spinor mass
$m = 1$, coupling constant $\lambda = 0.01$, cosmological constant
$\Lambda = 2/3$, $\ve_0 = 1$ and $\zeta = 2/3$.

\vskip 1 cm
\myfigures{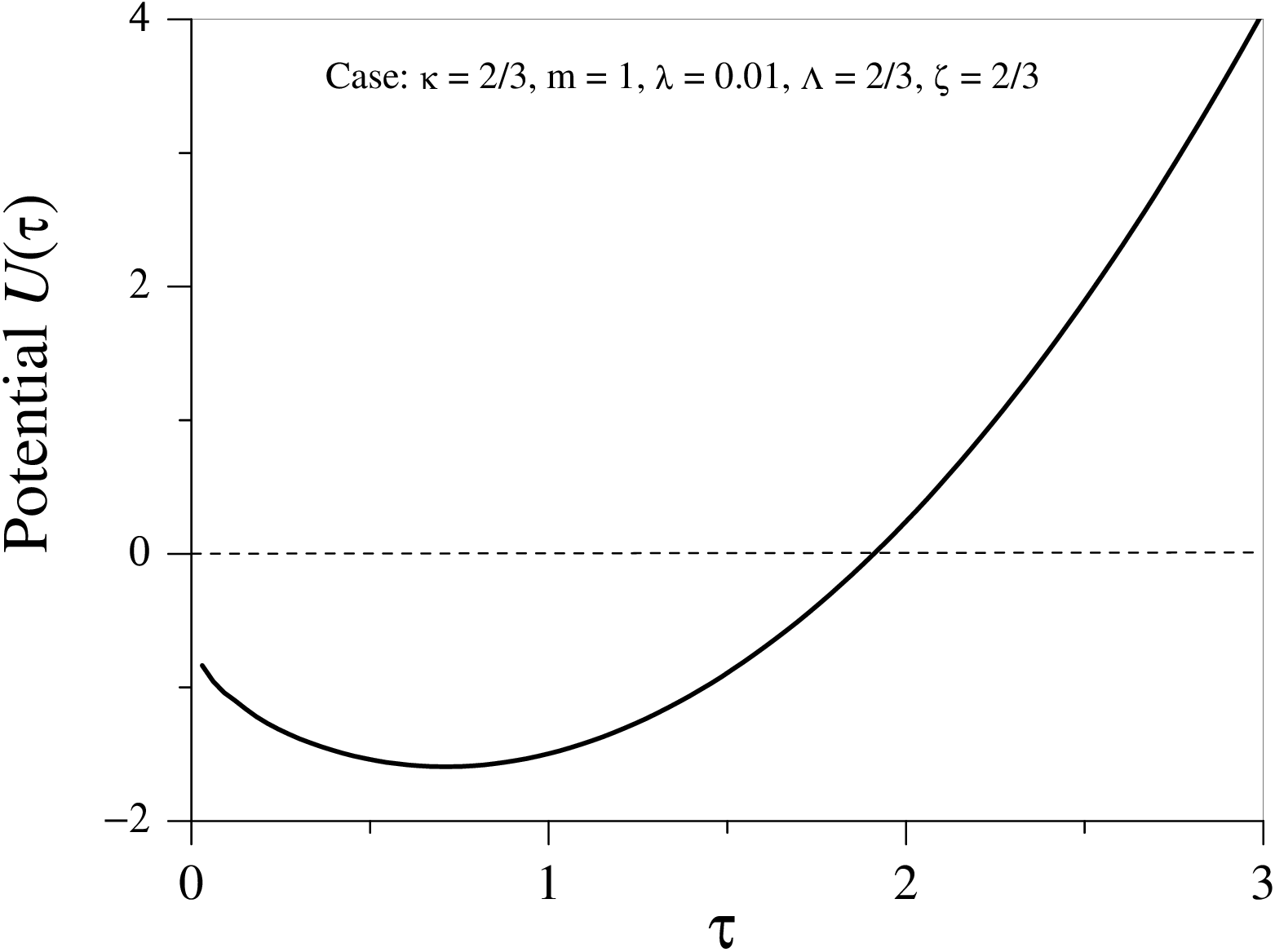}{0.45}{Perspective view of the potential 
$\mathcal U(\tau)$ with BI space-time being filled perfect fluid 
describing hard universe}{0.45}{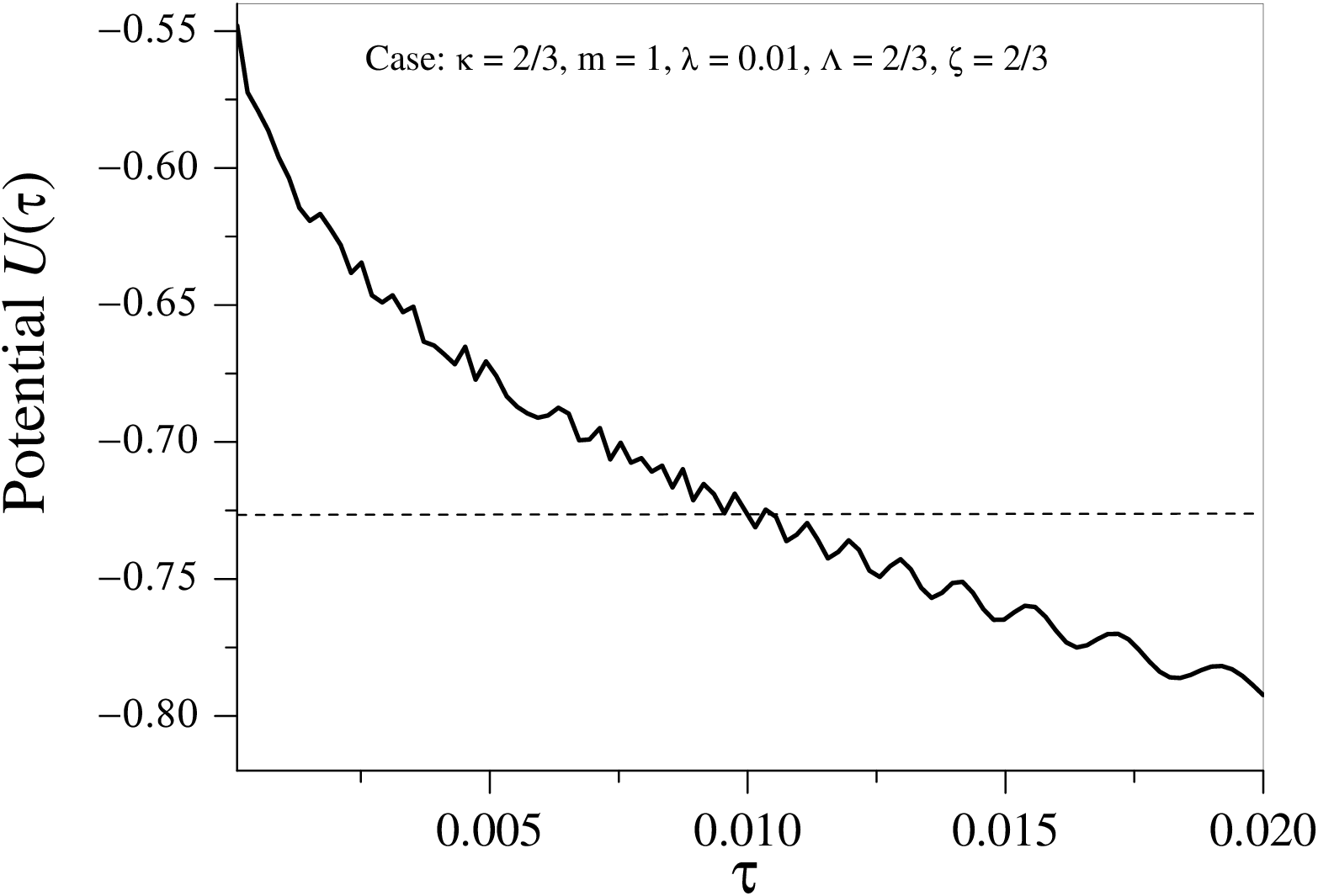}{0.50}{Extraordinary behavior of 
the potential in the vicinity of the singular point $\tau = 0$ that 
occurs due to the nonlinear term $F$.}{0.50}

It is clear from Fig.~\ref{c01} that an oscillatory mode of evolution
takes place, as was expected for a positive $\Lambda$. 

Let us now study the system for a negative $\Lambda$. Contrary to the
case with $F = S^n$, where all the solutions for a negative $\Lambda$
grow exponentially, in the case considered here an interesting
situation occurs for some special choice of parameters.

\vskip 1 cm
\myfigure{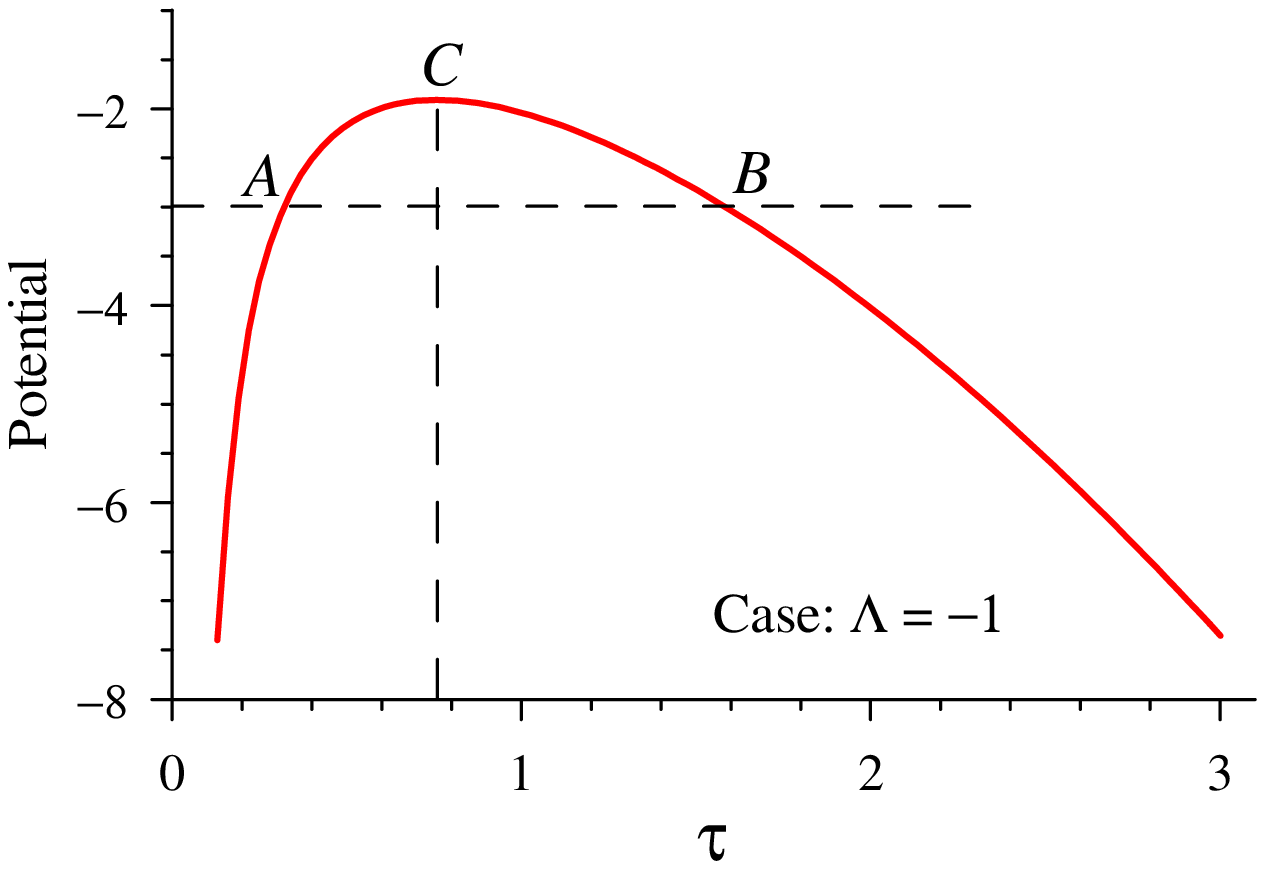}{0.75}{Perspective view of the potential 
$\mathcal U(\tau)$ with a negative $\Lambda$.}{0.75}

As one sees from Fig.~\ref{Bulbous}, depending on the integration
constant and initial value of $\tau$, the mode of evolution can be
both finite and exponential. For the integration constant being at the 
level $AB$ in Fig.~\ref{Bulbous} (here it is $-3$), with 
$\tau_0 \in (0, \tau_A)$ the evolution of $\tau$ finite and similar to
one in Fig.~\ref{tau6} corresponding to $E = 1$, whereas, for
$\tau_0 > \tau_B$ we have $\tau$ that is expanding exponentially. In case
of $E$ being at the same level with point $M$ we have the similar picture
of evolution, but in this case once $\tau$ reaches point $\tau_M$, 
the process of evolution would come to a halt. Thus we conclude that for a 
trigonometric interaction term the system even with a negative $\Lambda$ 
admits non-exponential mode of evolution. 

To investigate the dominant energy condition let us write the components 
of energy-momentum tensor. For simplicity we set $C_0 = 1$ and   
in term of $S$ for energy density we write
\begin{equation}
T_0^0 = m S + \frac{S^2}{2 (1 + \lambda \sin S)}
+ \ve_0 S^{1+\zeta}.
\label{edenp}
\end{equation}
Since $\tau$ is a positive quantity, $S$ is positive as well. 
As one sees from \eqref{edenp} for any positive value of $S$ and 
$\lambda < 1$ energy density is always positive definite and proportional
to $S^2$. Since $S = 1/\tau$, it means that $T_0^0$ has its maximum
as $\tau \to 0$ and tends to zero as $\tau \to \infty$.
 
On the For the pressure components we have
\begin{equation}
T_1^1 = T_2^2 = T_3^3 =
\frac{\lambda S^3 \cos S}{2 (1 + \lambda \sin S)^2} -
\frac{S^2}{2 ( 1 + \lambda \sin S)}  - \ve_0 \zeta S^{1+\zeta}.
\label{pdenp}
\end{equation}
As one sees, for a $\lambda < 1$ $T_1^1$ may both positive or negative
depending on the sign of $\cos S$. Moreover, its maximum value is 
proportional to $S^3$. Thus, in case of $F = \sin S$ for all possible values
of $\zeta$ and $\lambda$ (necessarily nontrivial) there exists intervals
$(S_n, S_{n+1})$ such that for $S \in (S_n, S_{n+1})$ the inequality
$T_0^0 < T_1^1$ takes place as it is shown in Fig.~\ref{Section}. 
Therefore we conclude that the regular solutions obtained in this case 
results in broken dominant energy condition. 

\vskip 1 cm
\myfigure{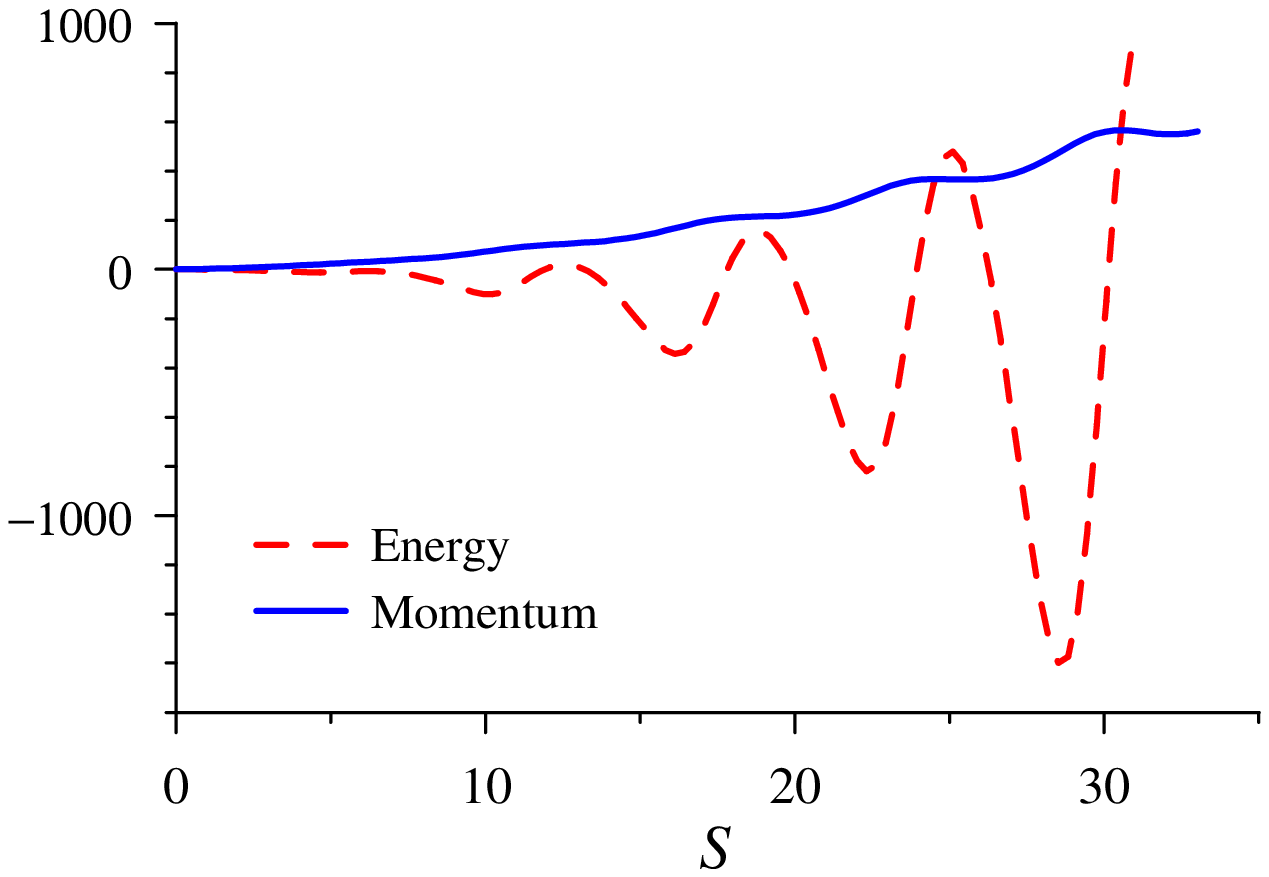}{0.75}{Perspective view of $T_0^0$ and $T_1^1$.}{0.75}

               \section{Conclusion}
Within the framework of the simplest model of interacting spinor 
and scalar fields it is shown that the $\Lambda$ term plays very 
important role in BI cosmology. In particular, it invokes 
oscillations in the model which is not the case when $\Lambda$ 
term remains absent. For a non-positive $\Lambda$ we find an universe
expanding exponentially, hence the initial anisotropy of the model
quickly dies away, whereas for a positive $\Lambda$ with the corresponding
choice of integration constant ${\cal E}$ one finds the oscillatory mode 
of expansion of the universe. In this case it is possible to construct
solutions those are always regular. It should be emphasized that if
the spinor field nonlinearity is generated by self-action the regularity
of the solutions obtained results in the violation of the dominant
energy condition of Penrose-Hawking theorem \cite{sahaprd}, whereas
in the case considered here, when the spinor field nonlinearity is induced
by the scalar one, regular solutions can be obtained even without breaking
the aforementioned condition. It should be noted that the dominant 
energy condition holds for $F$ being the power function of $I$ or $J$,
whereas it is not the case when $F$ is given as a trigonometric function
of its arguments. Note that in presence of $\Lambda$-term the role of other
parameters such as order of nonlinearity $n$, perfect fluid 
parameter $\zeta$ and spinor mass in the evolution process are 
rather local, while the global process are totally determined
by the $\Lambda$-term, e.g.,
for a positive $\Lambda$ we have always oscillatory mode, while for
a negative $\Lambda$ solutions are generally inflation-like though
for some special choices of problem parameters the oscillatory
mode of evolution can be attained. 

\begin{acknowledgments}
We would like to thank Prof. E.P. Zhidkov and G.N. Shikin for his kind attention to this work and helpful discussions.

\end{acknowledgments}


\end{document}